\documentclass{PoS}
\usepackage{latexsym}
\usepackage{graphicx}
\usepackage{amsmath}
\usepackage{amssymb}

\usepackage{url}

\title{Report on the 2019 Lattice Diversity and Inclusivity Survey}

\ShortTitle{Diversity and Inclusivity Report}

\author{\speaker{Lattice Diversity and Inclusivity Committee}}
\author{Christopher Aubin%
        \\
       Department of Physics and Engineering Physics, 
       Fordham University, Bronx, NY 10458, USA\\
       E-mail: \email{caubin@fordham.edu}
       }

\author{Gunnar Bali\\
Institut f\"ur Theoretische Physik, Universit\"at Regensburg, 
93040 Regensburg, Germany\\
E-mail: \email{gunnar.bali@ur.de}
}
\author{Luigi Del Debbio\\
The Higgs Centre for Theoretical Physics, 
University of Edinburgh, 
Edinburgh EH9 3FD, UK\\
E-mail: \email{luigi.del.debbio@ed.ac.uk}
}
\author{William Detmold\\
Center for Theoretical Physics, Massachusetts Institute of 
Technology, Cambridge, MA 02139, USA\\
E-mail: \email{wdetmold@mit.edu}
}
\author{Vera G\"ulpers\\
School of Physics and Astronomy, 
University of Edinburgh, 
Edinburgh EH9 3JZ, UK\\
E-mail: \email{Vera.Guelpers@ed.ac.uk}
}
\author{Sophie Hollitt\\
Fakult\"at Physik, Technische 
Universit\"at Dortmund, 44221 
Dortmund, Germany\\
E-mail: \email{sophie.hollitt@tu-dortmund.de}
}
\author{Huey-Wen Lin\\
Department of Physics and Astronomy, 
Michigan State University, East Lansig, MI 48824, USA\\
E-mail: \email{hwlin@pa.msu.edu}
}
\author{Liuming Liu\\
Institute of Modern Physics,
Chinese Academy of Sciences, Lanzhou 730000, 
China\\
E-mail: \email{liuming@impcas.ac.cn}
}
\author{Sin\'ead M.\ Ryan\\
School of Mathematics, Trinity College
Dublin 2, Ireland\\
E-mail: \email{ryan@maths.tcd.ie}
}

\abstract{}

\FullConference{37th International Symposium on Lattice Field Theory - Lattice2019\\
		16-22 June 2019\\
		Wuhan, China}

\begin{document}

\section{Overview}

In 2018, the International Advisory Committee for the 
2019 Lattice conference supported a proposal to form a 
Diversity and Inclusivity committee. The elected members 
of this committee were then charged with writing a 
code of conduct -- which all participants agreed to 
as part of the conference registration; providing 
guidelines for conference session chairs; and lastly 
conducting a survey to assess diversity and inclusivity 
in the Lattice community and presenting the results 
in the conference poster session and in these proceedings. 

The survey was undertaken using a 
Google form (and a China-based equivalent) and invitations 
were emailed to the attendee list from Lattice 2018 and to 
the \href{mailto::latticenews-l@list.indiana.edu}{latticenews mailing list}. 
The poll was open for 
approximately one month and a follow-up invitation was 
sent with a week to go. All questions included a free-response
box for answers not included in the multiple-choice 
options as well as a ``Prefer not to answer.''

There were 174 responses to the survey, many of which provided 
very thoughtful input. Half of the responses came from Europe, 
1/3 from the US and the remaining 1/6 from Asia/Oceania/Africa.
 
We find that the 
field is not particularly diverse and is (unsurprisingly) 
dominated by 
Caucasian/white heterosexual males. The proportion of male 
respondents was 83\%, of female respondents was 11\%, and 
3\% identified as gender fluid/non-binary. The percentage 
of female respondents is consistent with the average rate 
of participation at recent lattice conferences. The rate of 
transgender identification was 2\%. The rate of respondents 
identifying as gay (4\%) is within the range of expectations 
based on similar surveys \cite{aps,lattice2016,beckfordichep,yacoobichep,PhysRev}.
 
There are issues with inclusivity in the community; 25-30\% 
of respondents indicated not feeling comfortable 
infrequently or more often at 
talks, social events and even conference accommodations. 
A significant 
number of respondents (about 35\%) has observed or had negative 
experiences. The nature of these experiences was not determined 
in the survey.
 
Histograms of responses to survey questions based on 
different demographic data collected are shown below in 
Sec.~\ref{sec:correlations}. For the free text 
questions, results are summarised here but explicit
responses are not included in this write-up.
 
Respondents felt the gender balance of plenary speakers has 
improved. This impression is not supported by the statistics 
of the last ten conferences.
Respondents felt that the percentage of women in the field 
was higher than it actually is.
Respondents are generally supportive of efforts to improve 
diversity (e.g., a diversity committee, 2018 lunch talks).
Respondents felt the tone of discussions has become less 
aggressive in comparison to the 90's.
Future surveys should probe more detailed reasons for 
the discomfort or negative experiences of the respondents.
 
In these proceedings we discuss recommendations
based on the results of the survey in Sec.~\ref{sec:recs}. 
In Sec.~\ref{sec:correlations} we show the results from the
questions regarding comfort levels and negative experiences
from the survey, and we briefly conclude in Sec.~\ref{sec:conc}.
We include the responses to the demographics questions in Appendix
1 and an update of the Women in Lattice statistics in Appendix 2.

\section{Recommendations}\label{sec:recs}

After reviewing the results from the survey, the committee
has the following recommendations for the lattice community.

\begin{enumerate}
\item Formalize the diversity committee and maintain the 
code of conduct. Survey diversity periodically, updating 
the questionnaire based on the lessons learned
from this survey and keep statistical records.

\item Continue the ``Women in Lattice'' lunches, opening this
event to any interested participants.

\item Regularly hold a plenary on the topic of diversity and 
inclusivity.

\item Further increase the level of diversity in 
plenary talks and 
session chairs (not just monitoring the gender balance).

\item Provide financial support for minorities as well as for students.

\item Foster mentoring of junior colleagues. Identify senior 
physicists to meet with a group of younger scientists for career discussions.

\item Attempt to promote international mixing between 
participants at different career stages.

\item Establish a clear mechanism for reporting and responding 
to incidents of exclusion and harassment.

\item Develop an informal practical guide for 
future local organizing committees, related 
to diversity and inclusivity, 
to help ensure a positive atmosphere 
for all participants. This should be updated and passed on from
year to year, along with the lattice conference manual. 
\end{enumerate}

\section{Summary of the Survey Results}\label{sec:correlations}

The following histograms correlate the questions ``Did you feel 
comfortable at the most recent lattice conference you attended?'' 
and ``Have you had direct or indirect (personally or observed) 
negative experiences at a lattice conference?'' with various 
demographics questions. We have averaged each 
person's responses over 
the different parts of the conference 
that were asked about in the 
survey (such as ``during scientific 
talks'' or ``during social events'') 
for simplicity. Additionally, we 
have normalized each response by the 
total number of respondents in that category (such 
as ``male'' or ``female'') 
for easier comparison. In all of the histograms below,
we have made the following abbreviations for the
labels
\begin{center}
	\textbf{A:} Always\qquad \qquad 
	\textbf{F:} Frequently\qquad \qquad 
	\textbf{S:} Sometimes

	\textbf{I:} Infrequently\qquad \qquad 
	\textbf{N:} Never\qquad \qquad 
	\textbf{N/A:} Not applicable
\end{center}

\subsection{Age}
 
\begin{figure}[htbp]
\begin{center}
\includegraphics[width=2.9in]{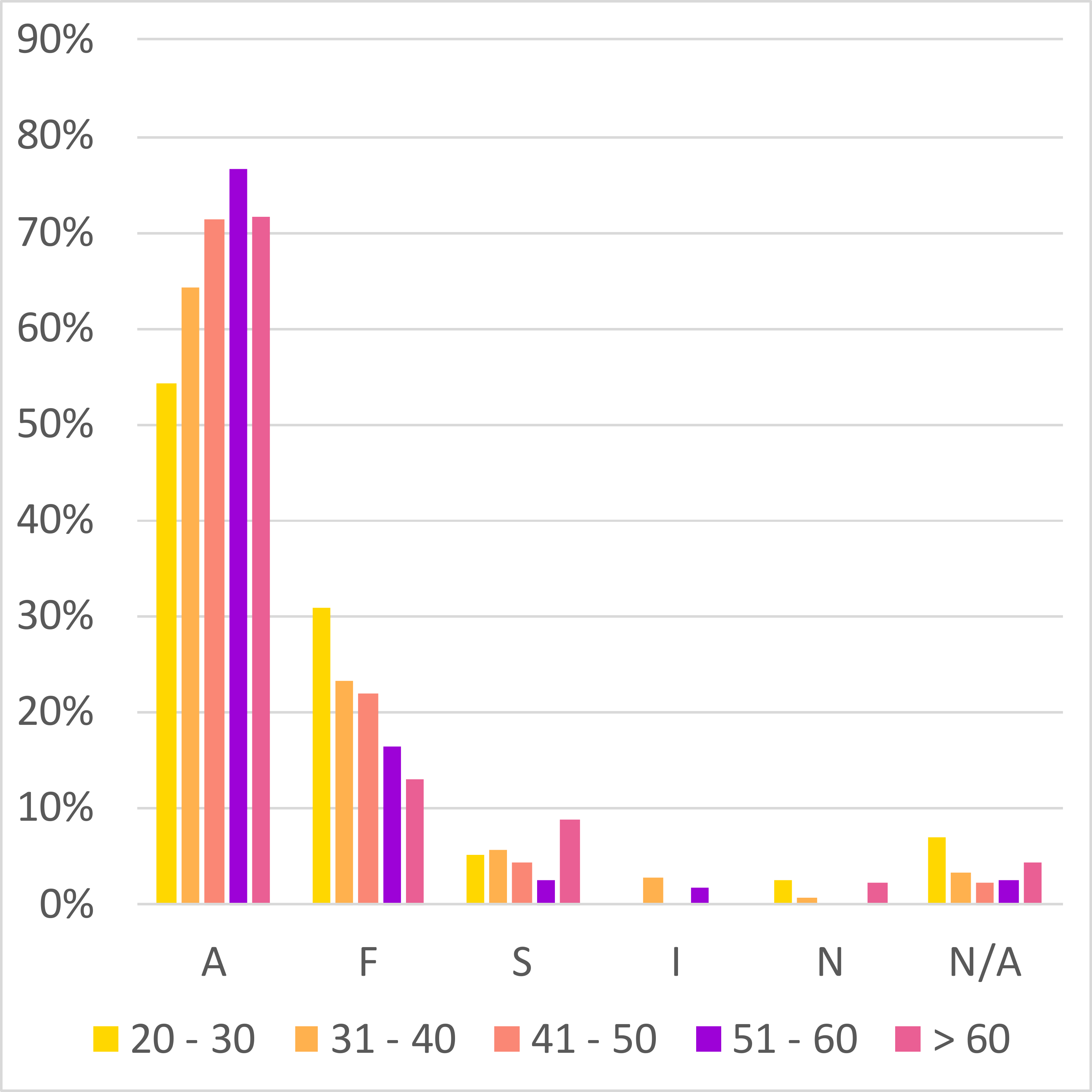}
\includegraphics[width=2.9in]{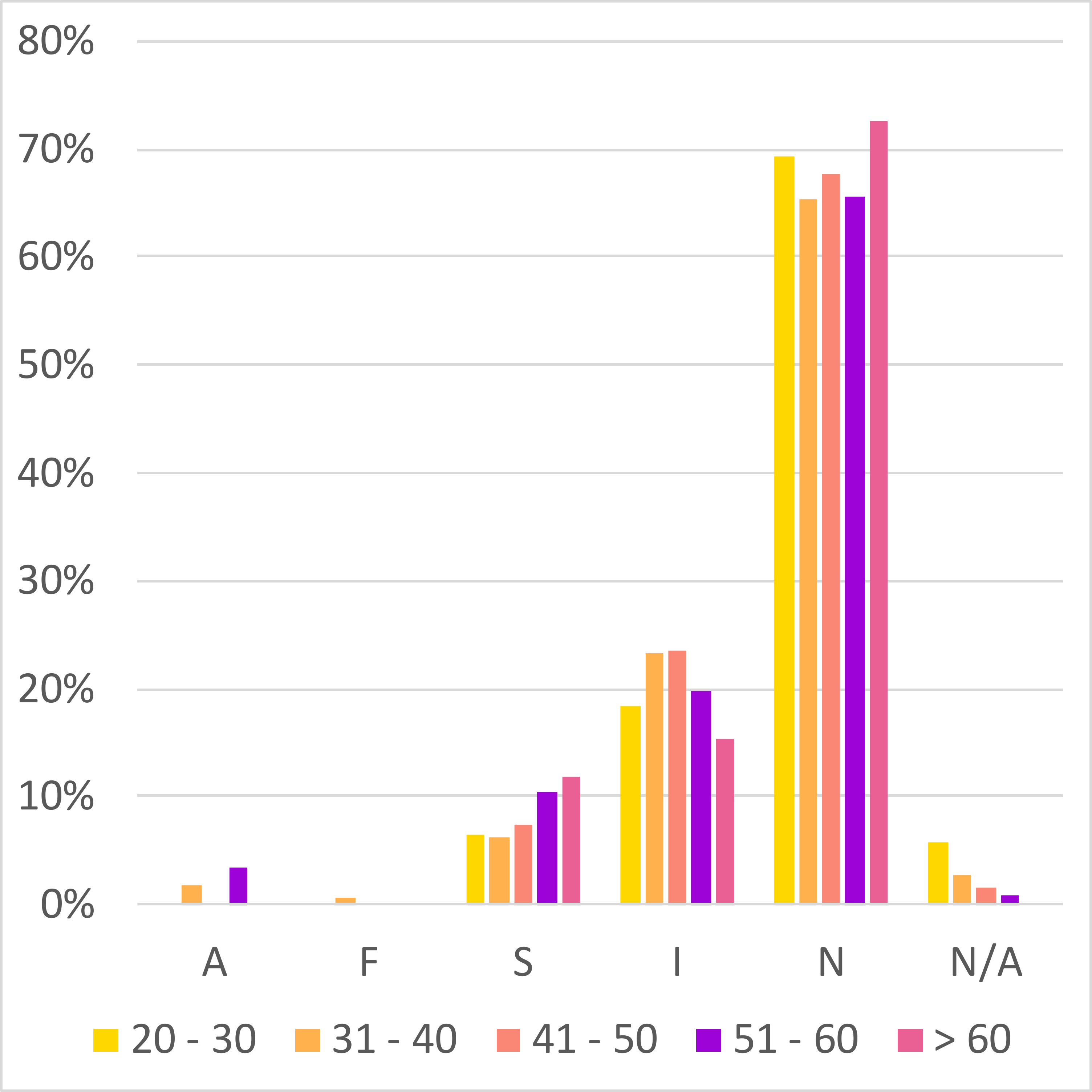}

(a)\hspace{3in}(b)

\caption{Answers to the following questions by age: (a) ``Did you feel 
comfortable at the most recent lattice conference you attended?''
(b) ``Have you had direct or indirect (personally or observed) 
negative experiences at a lattice conference?''}
\label{fig:comfortnegage}
\end{center}
\end{figure}

Given the few responses in the age category ``$>$ 70'' we have combined 
those with the 61 - 70 age group. We show in Fig.~\ref{fig:comfortnegage} 
 the answers to these questions by age.

\subsection{Rank/Position}

\begin{figure}[htbp]
\begin{center}
\includegraphics[width=2.9in]{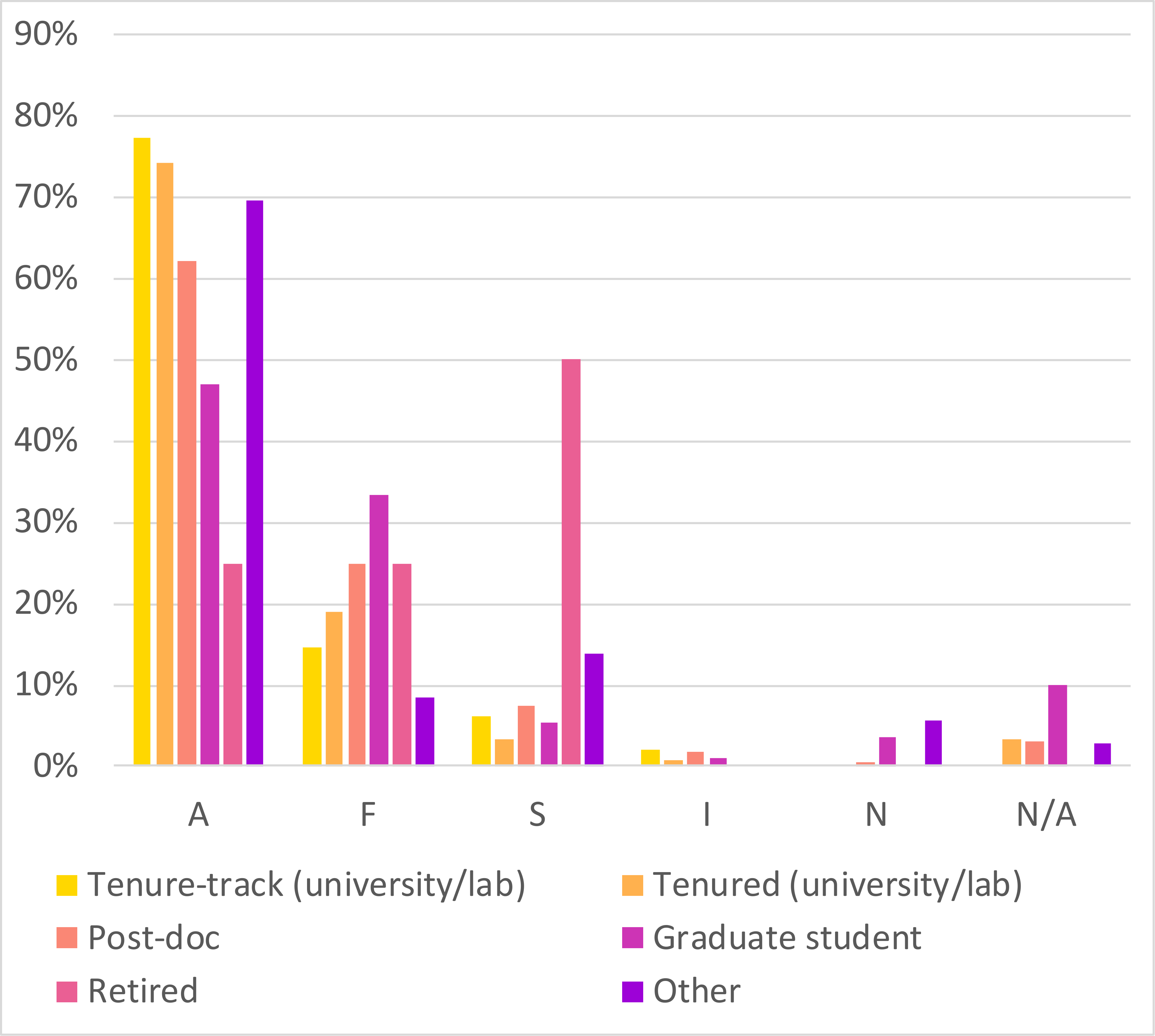}
\includegraphics[width=2.9in]{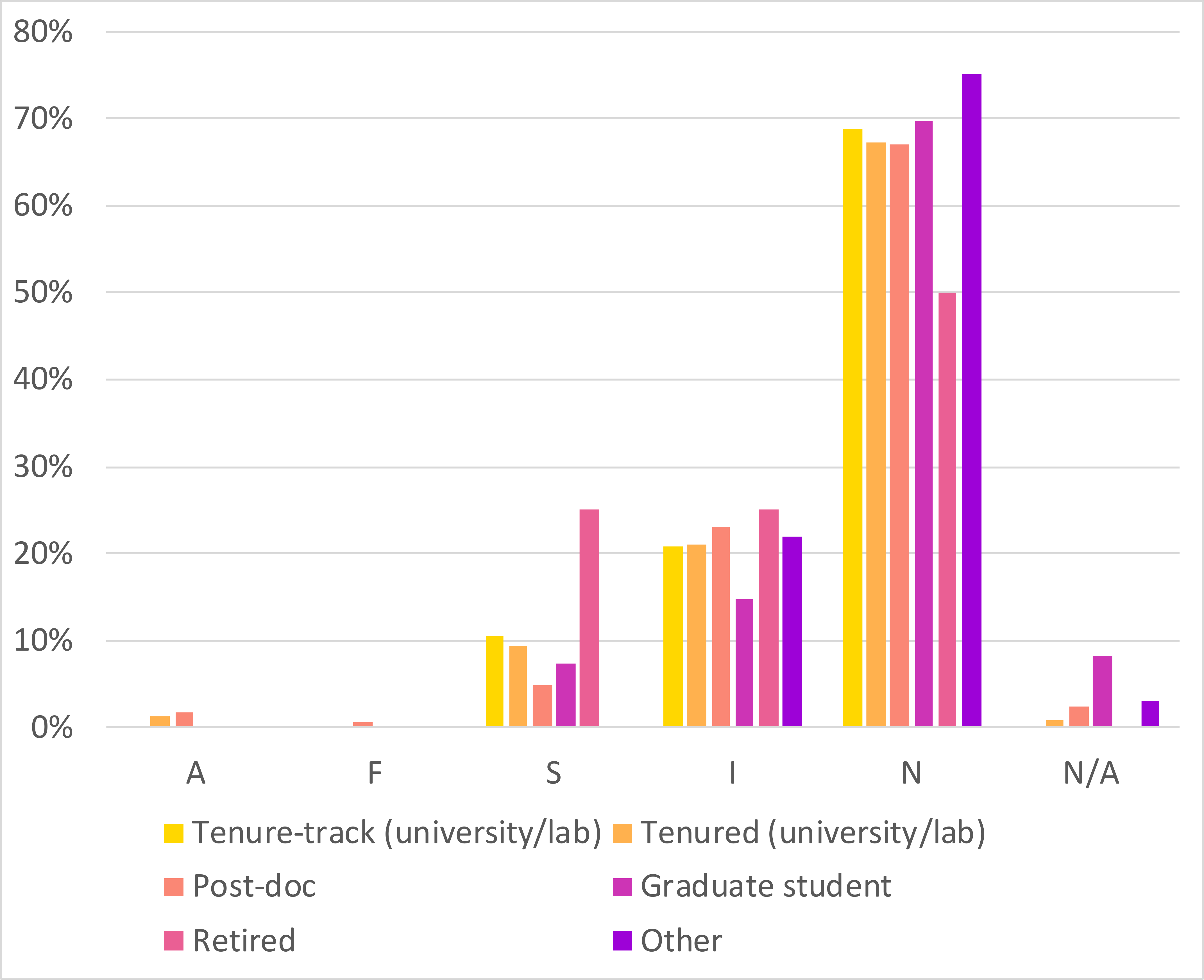}

(a)\hspace{3in}(b)

\caption{Answers to the following questions by rank: (a) ``Did you feel 
comfortable at the most recent lattice conference you attended?''
(b) ``Have you had direct or indirect (personally or observed) 
negative experiences at a lattice conference?''}
\label{fig:comfortnegrank}
\end{center}
\end{figure}

Figure~\ref{fig:comfortnegrank} presents the responses 
to the academic climate questions with respect to academic rank. 
In these categories the ``Other'' category 
includes the responses: 
Industry, Editor, Adjunct, Untenured faculty, Former Graduate 
Student, and those who selected ``Prefer not to answer.'' Here 
it seems as though lack of comfort/more negative experiences 
are seen by those in the ``Retired'' group, and to a lesser 
extent the ``Graduate Students.''

\subsection{Gender Identity}
 
\begin{figure}[htbp]
\begin{center}
\includegraphics[width=2.9in]{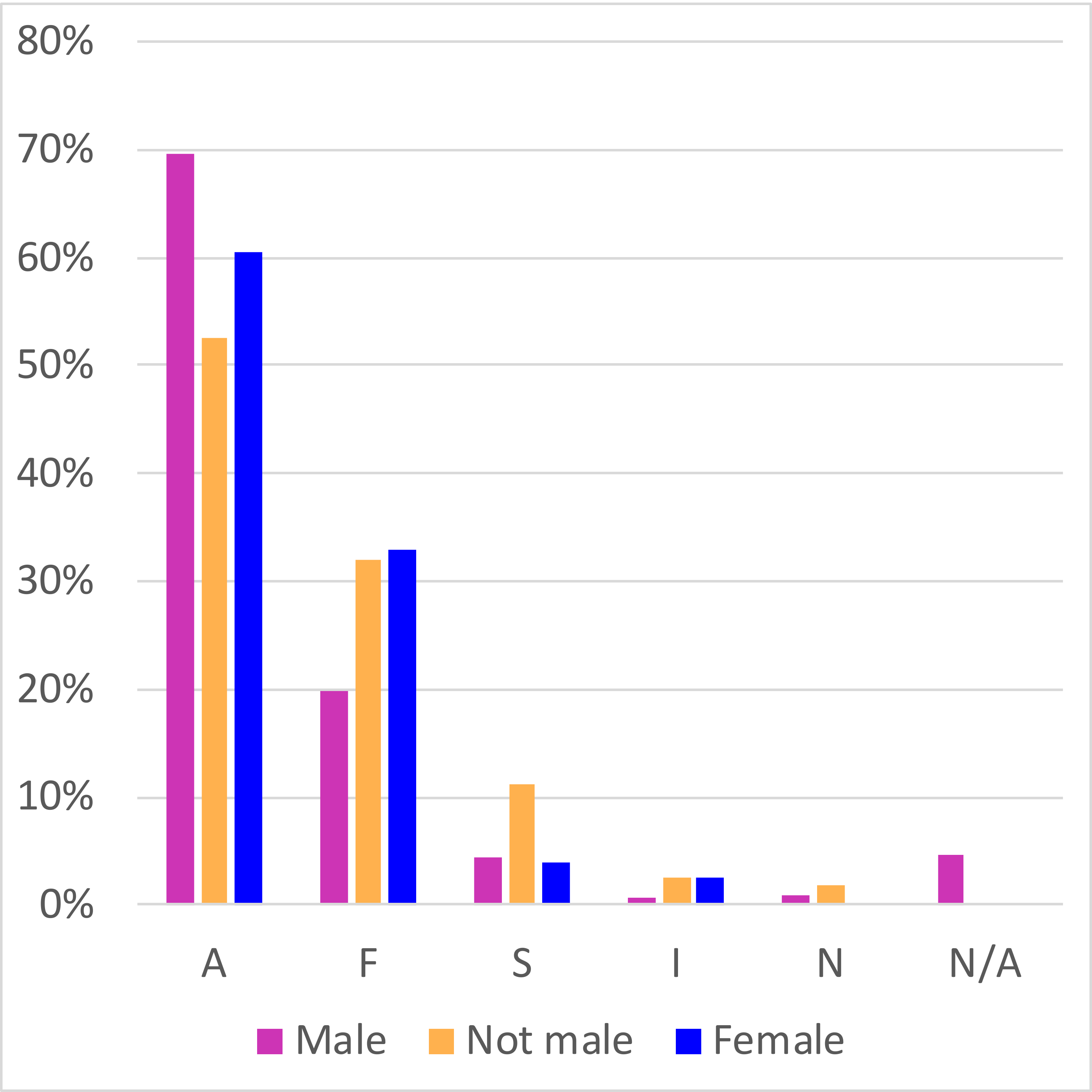}
\includegraphics[width=2.9in]{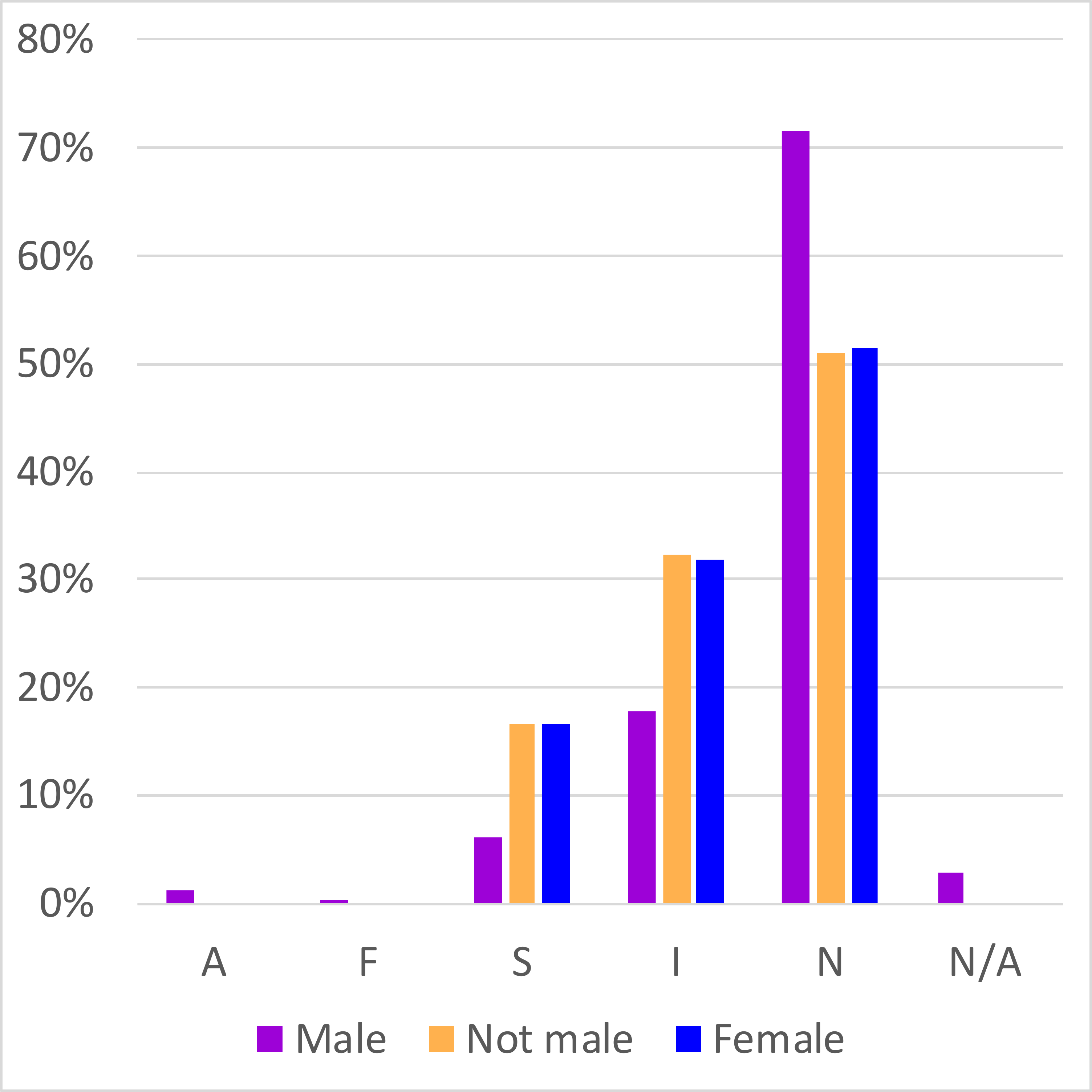}

(a)\hspace{3in}(b)

\caption{Answers to the following questions by gender 
identity: (a) ``Did you feel 
comfortable at the most recent lattice conference you attended?''
(b) ``Have you had direct or indirect (personally or observed) 
negative experiences at a lattice conference?''}
\label{fig:comfortneggender}
\end{center}
\end{figure}

For the plots in Fig.~\ref{fig:comfortneggender}, 
we make three categories in order to 
not just compare responses between those who identify as male 
or female but also those who do not identify as 
male. This latter category includes all responses that 
were not male: female, gender fluid/non-binary, write-in 
responses and ``prefer not to answer.'' It is clear that 
negative experiences are more common for those who do not 
identify as male.

\subsection{LGBTQIA Responses}
 
\begin{figure}[htbp]
\begin{center}
\includegraphics[width=2.9in]{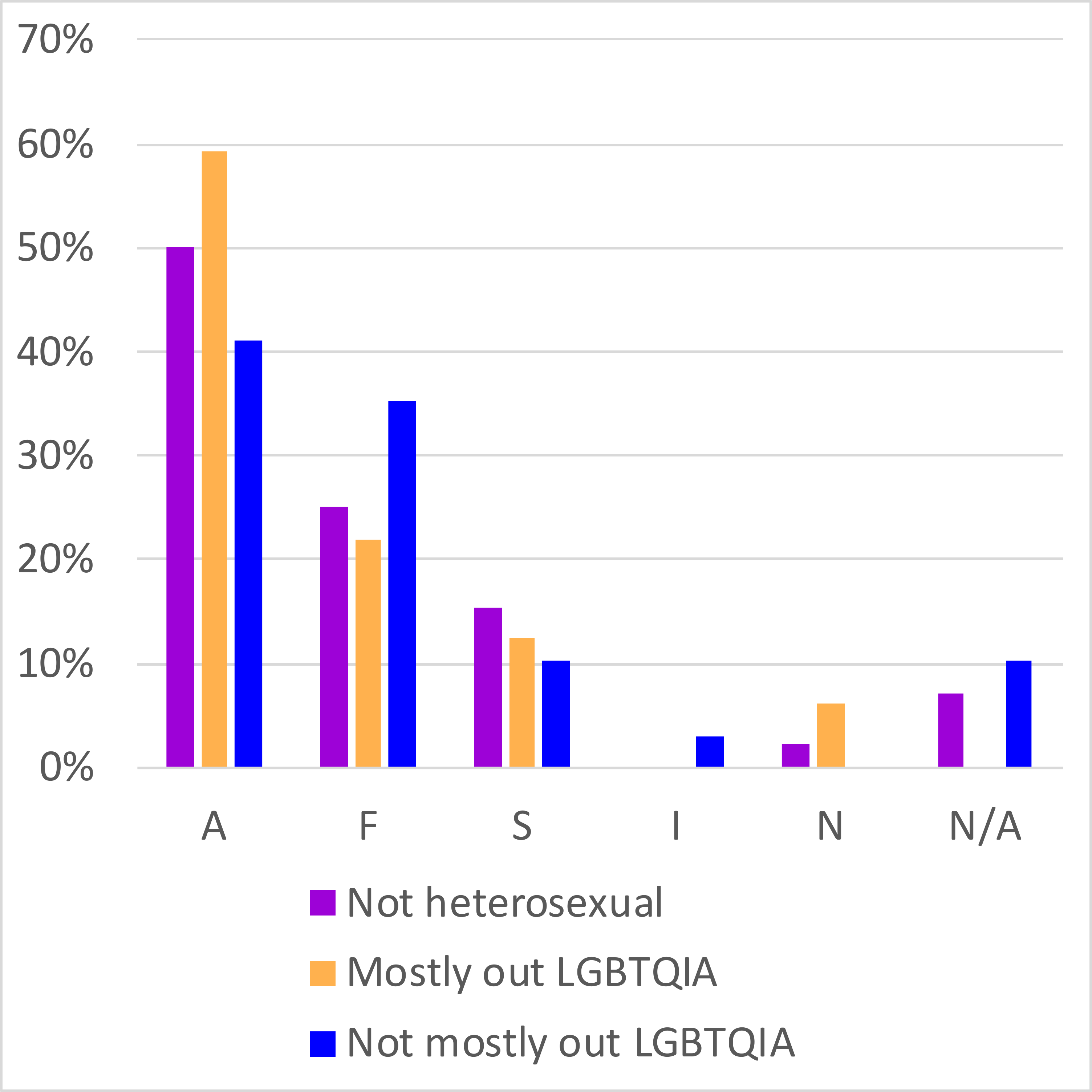}
\includegraphics[width=2.9in]{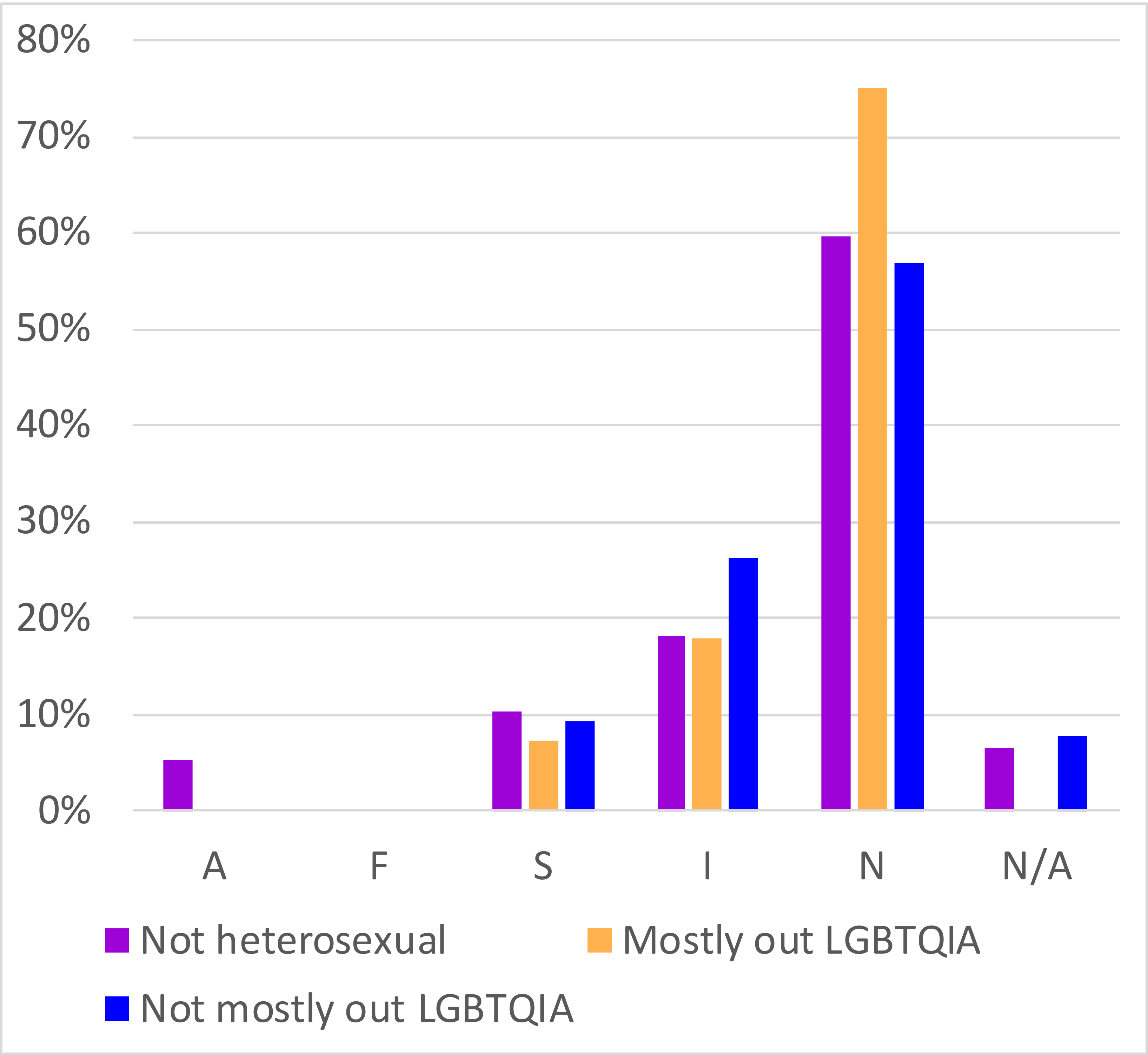}

(a)\hspace{3in}(b)

\caption{Answers to the following questions by LGBTQIA 
identity: (a) ``Did you feel 
comfortable at the most recent lattice conference you attended?''
(b) ``Have you had direct or indirect (personally or observed) 
negative experiences at a lattice conference?''}
\label{fig:comfortnegLGBTQIA}
\end{center}
\end{figure}

In Fig.~\ref{fig:comfortnegLGBTQIA}, 
we look at responses from those who do not identify as 
heterosexual as well as those with any 
LGBTQIA (Lesbian-Gay-Bisexual-Trans-Queer-Intersex-Asexual) identity 
(broken up into those who are mostly ``out'' to the lattice community 
and those who are not). Those who are out are less likely 
to have felt uncomfortable or to have had negative experiences.

\begin{figure}[htbp]
\begin{center}
\includegraphics[width=2.9in]{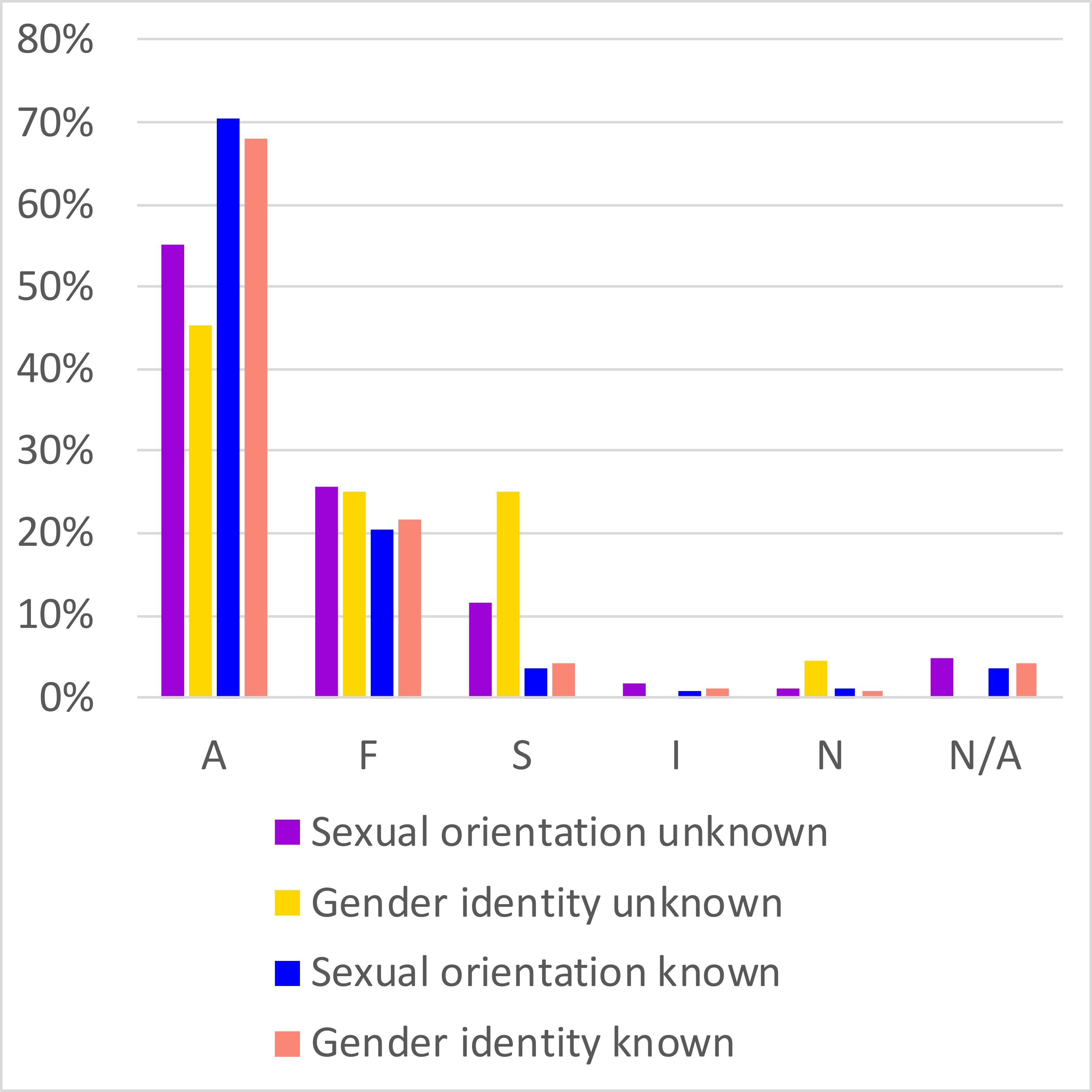}
\includegraphics[width=2.9in]{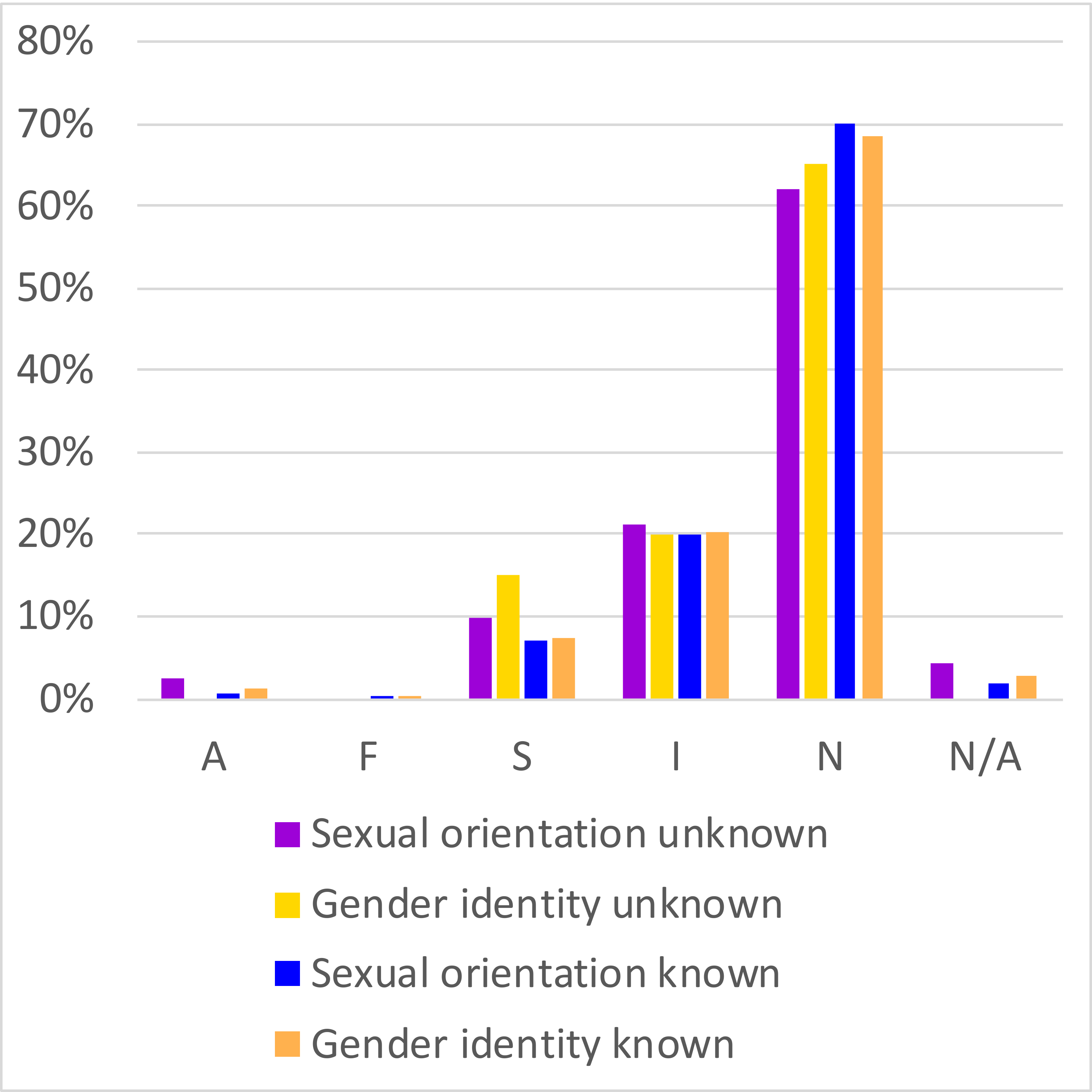}

(a)\hspace{3in}(b)

\caption{Answers to the following questions by perceived knowledge of LGBTQIA
or gender identity: (a) ``Did you feel 
comfortable at the most recent lattice conference you attended?''
(b) ``Have you had direct or indirect (personally or observed) 
negative experiences at a lattice conference?''}
\label{fig:comfortnegunknown}
\end{center}
\end{figure}

Additionally in Fig.~\ref{fig:comfortnegunknown}, 
we examine the responses from those who think 
that their identity (be it gender identity or sexual 
orientation) is known by others in the lattice community. 
Those whose identities are known appear to feel more 
comfortable at the lattice conferences and are slightly 
less likely to have had negative experiences. This makes 
it clear that we should work on having a more positive 
and inclusive atmosphere.

\subsection{Comfort Level and negative experiences during conferences}

In Table~\ref{tab:comfort}(a), we show the raw numbers from responses
to the question ``Did you feel 
comfortable at the most recent lattice conference you attended?'' 
For this table, 1 is ``never comfortable'' and 5 is ``always comfortable''
at various times during the conference. (In both tables, ``Accomm.'' stands
for Accommodations.)

\begin{table}[hbtp]
\begin{center}
\begin{tabular}{ccccc}
& Talks & Events & Breaks & Accom.\\
\hline
\hline
1 & 2 & 2 & 2 & 0\\
\hline
2 & 2 & 2 & 2 & 0\\
\hline
3 & 6 & 13 & 13 & 6\\
\hline
4 & 33 & 40 & 43 & 34\\
\hline
5 & 125 & 111 & 108 & 114\\
\hline
n/a & 4 & 4 & 4 & 15\\
\hline
\end{tabular}\hfill
\begin{tabular}{ccccc}
& Talks & Events & Breaks & Accom.\\
\hline
\hline
1 & 95 & 113 & 118 & 136\\
\hline
2 & 48 & 32 & 34 & 23\\
\hline
3 & 21 & 19 & 12 & 0\\
\hline
4 & 0 & 0 & 0 & 0\\
\hline
5 & 2 & 0 & 2 & 2\\
\hline
n/a & 3 & 3 & 3 & 8\\
\hline
\end{tabular}

(a)\hspace{3in}(b)

\caption{In (a), 1 is ``never comfortable'' and 5 is 
``always comfortable'' at various times during the conference. In (b),
1 is ``never had a negative experience.'' 5 is 
``always had a negative experience.''}
\label{tab:comfort}
\end{center}
\end{table}
The majority of respondents always felt comfortable. 
However it is striking that about 25\% of respondents felt 
uncomfortable even some of the time during talks, social events 
and even at conference accommodations. The highest levels 
of discomfort were at social events and informal gatherings/breaks.

In Table~\ref{tab:comfort}(b) we show the raw numbers from responses
for the question
``Have you had direct or indirect (personally or observed) 
negative experiences at a lattice conference?'' Here, 
1 is ``never had a negative experience.'' 5 is 
``always had a negative experience.'' 
In each case the majority did not report negative experiences, however
about 35\% of respondents have been subject to, or 
witnessed, negative experiences.

\section{Conclusion}\label{sec:conc}

As discussed above, in general the community is amenable to 
programs such as this committee and other ways to
improve diversity in the field. Steps should be taken
in order to (as much as possible) improve inclusivity
for all participants at all conferences.

The IAC has approved our proposal to form a 
standing Diversity
and Inclusivity Committee. This Committee will 
update the Code of
Conduct, promote diversity and inclusivity in 
the community and periodically survey the progress made. We ask that any suggestions
for improvement are forwarded to the members of this Committee.

\section*{Appendix 1: Demographics Questions}\label{app1}

In this appendix we include graphs showing the direct answers to 
the demographics questions in the survey. 

\begin{figure}[htbp]
\begin{center}
\includegraphics[width=2.5in]{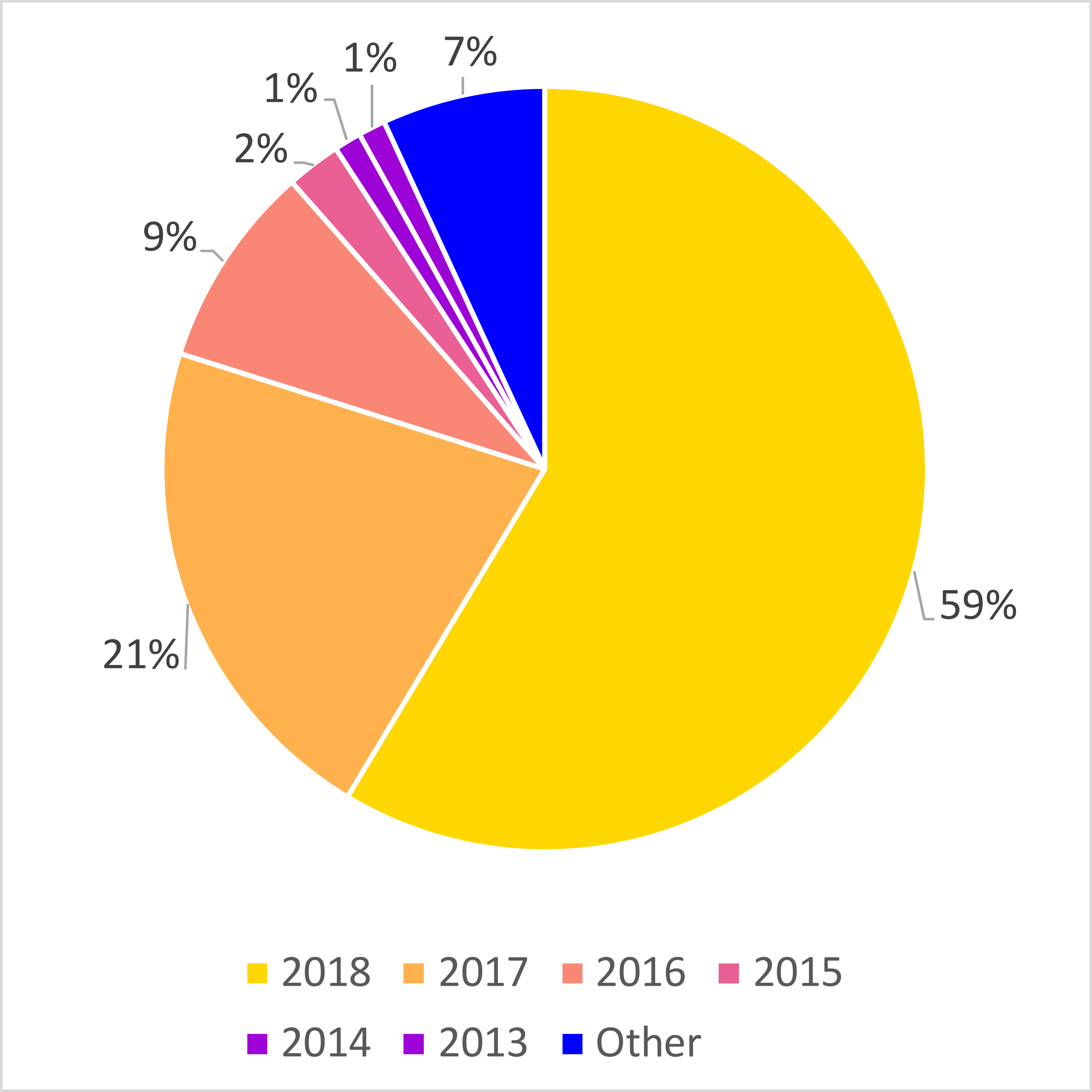}
\hspace{.8in}
\includegraphics[width=2.5in]{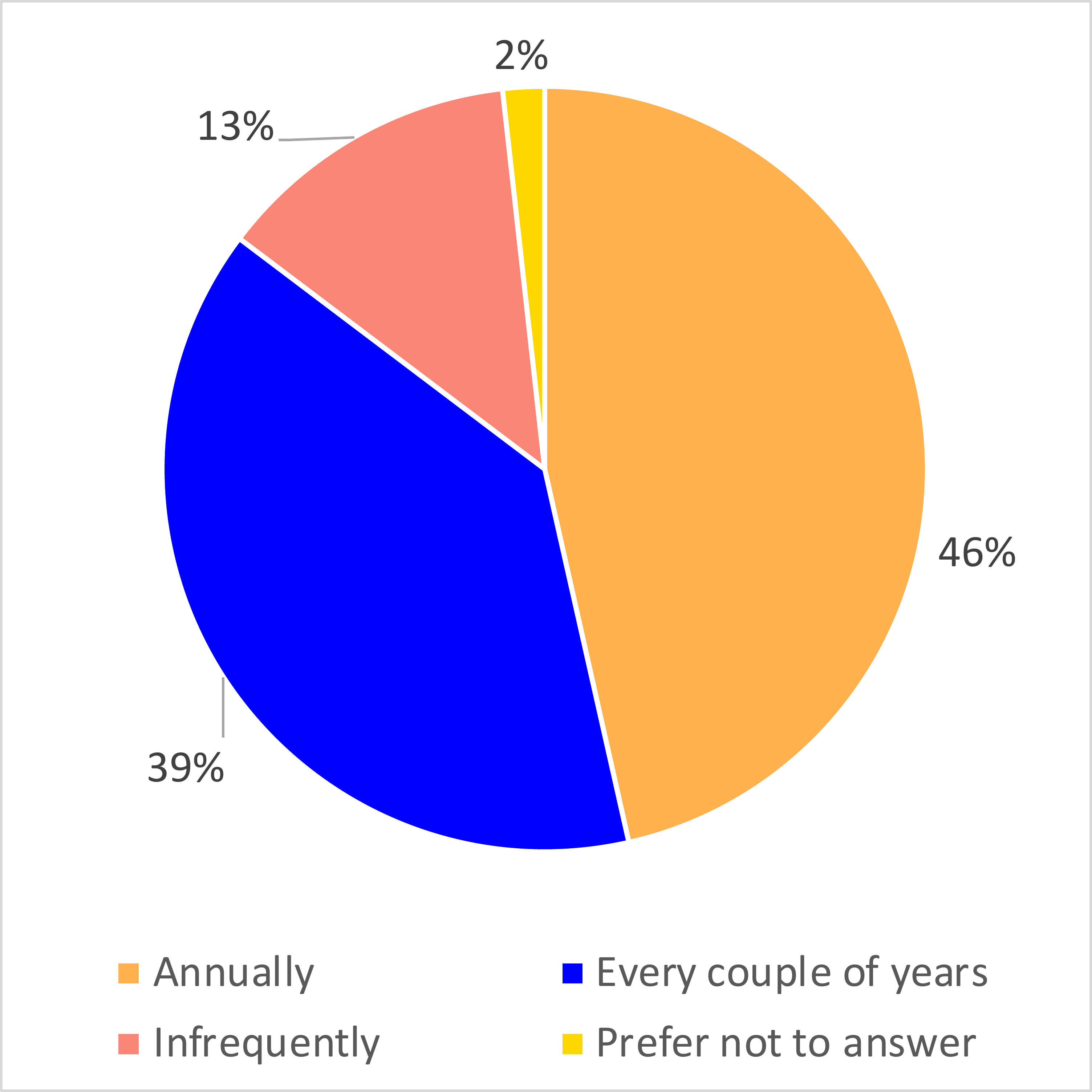}

(a)\hspace{3in}(b)

\caption{(a) What year was the last lattice 
conference you attended? 
(b) How often do you attend the lattice
conference?}
\label{fig:latticeattendance}
\end{center}
\end{figure}

\begin{figure}[htbp]
\begin{center}
\includegraphics[width=2.5in]{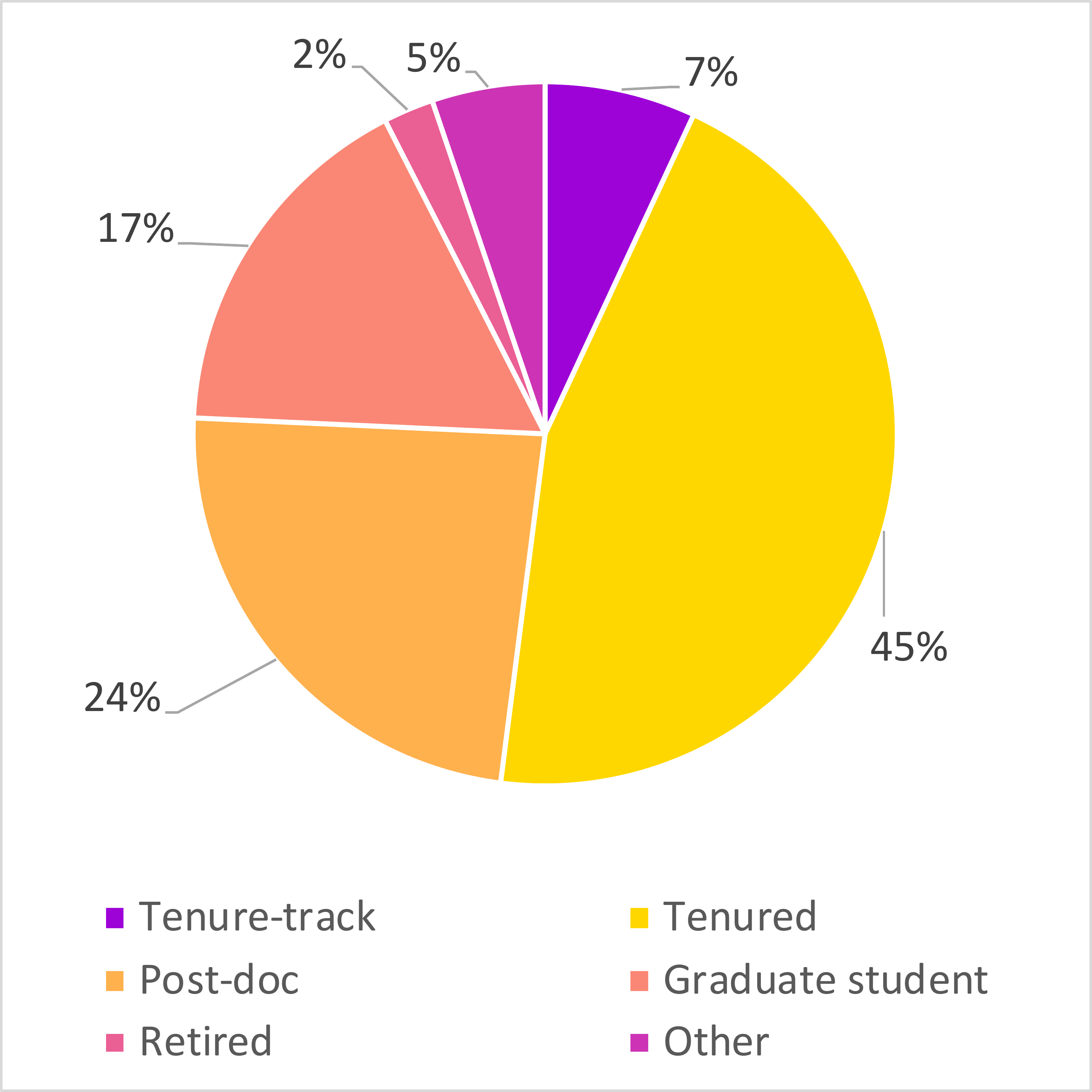}
\hspace{.8in}
\includegraphics[width=2.5in]{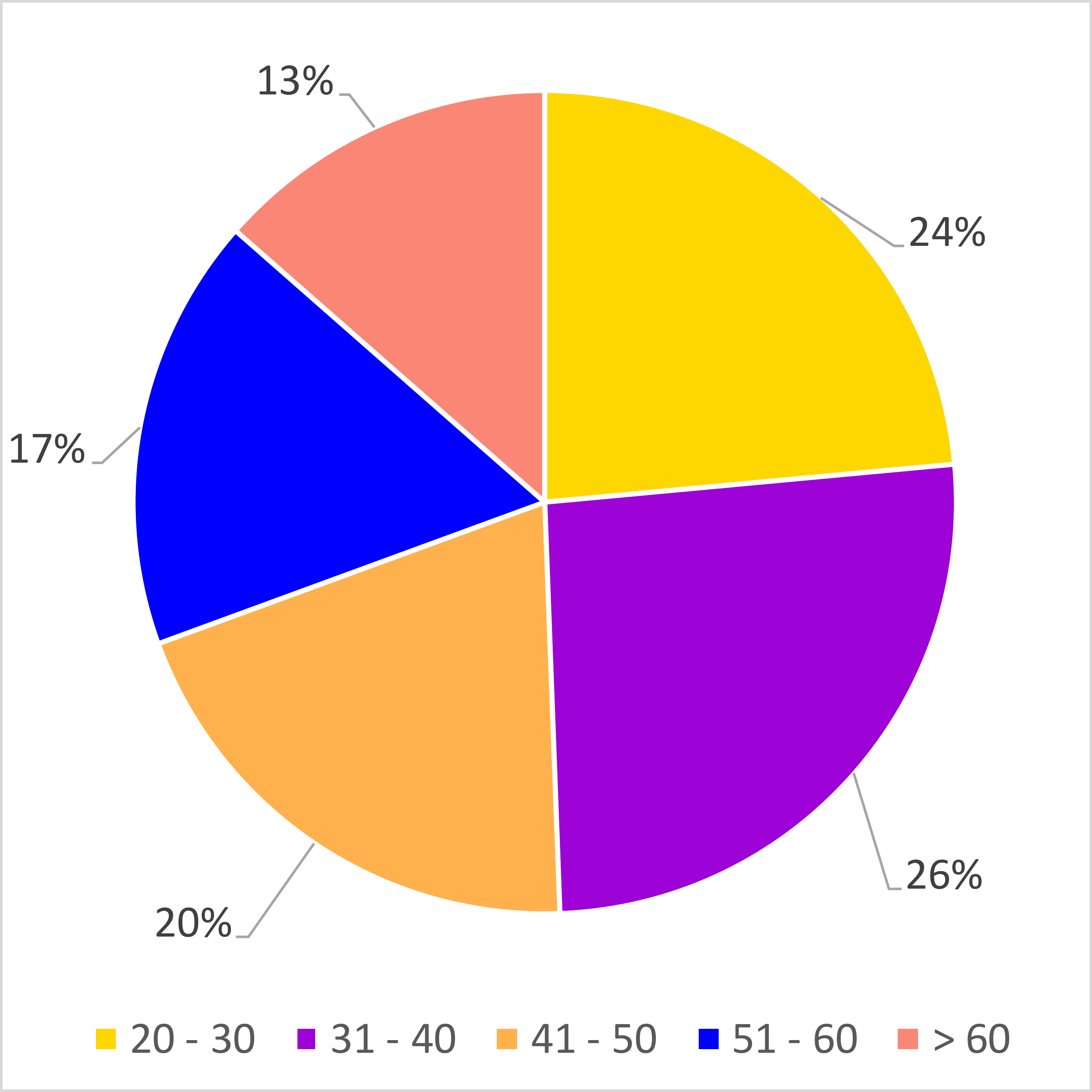}

(a)\hspace{3in}(b)

\caption{(a) What is your current rank? (b) What is your age?}
\label{fig:rankage}
\end{center}
\end{figure}

\begin{figure}[htbp]
\begin{center}
\includegraphics[width=2.5in]{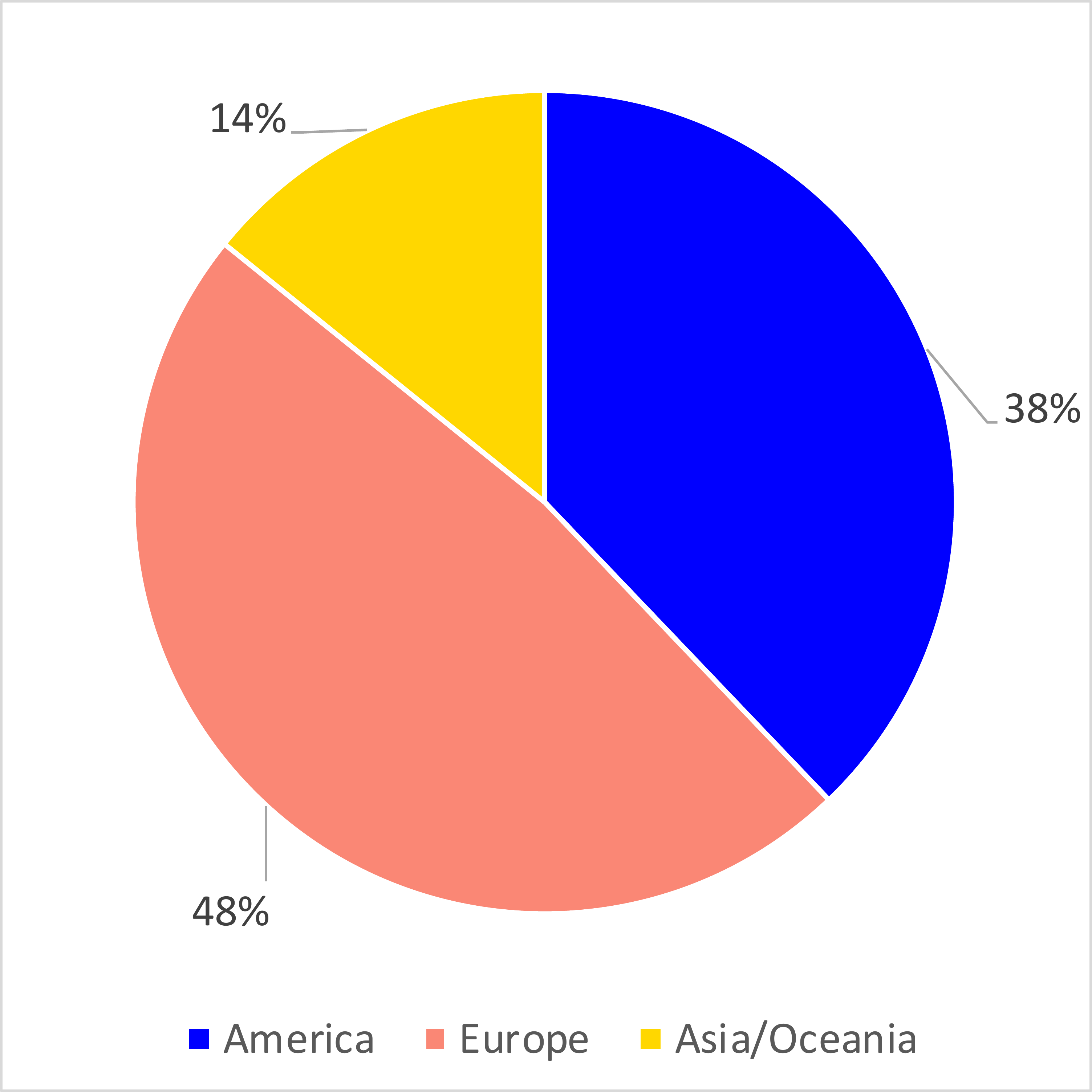}
\hspace{.8in}
\includegraphics[width=2.5in]{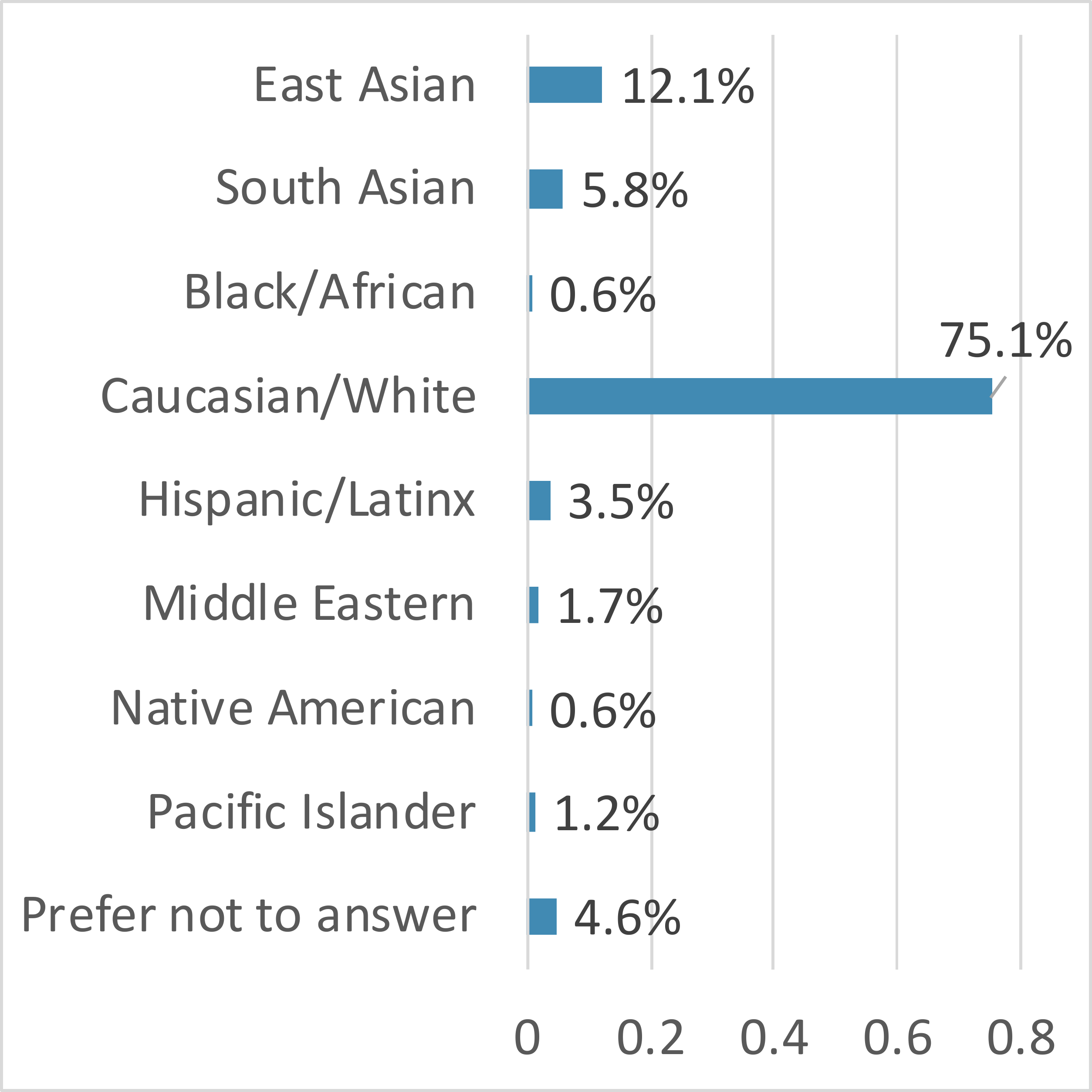}

(a)\hspace{3in}(b)

\caption{(a) What geographic region do you work in?
(b) What is your ethnicity?}
\label{fig:regionethnicity}
\end{center}
\end{figure}

\begin{figure}[htbp]
\begin{center}
\includegraphics[width=2.5in]{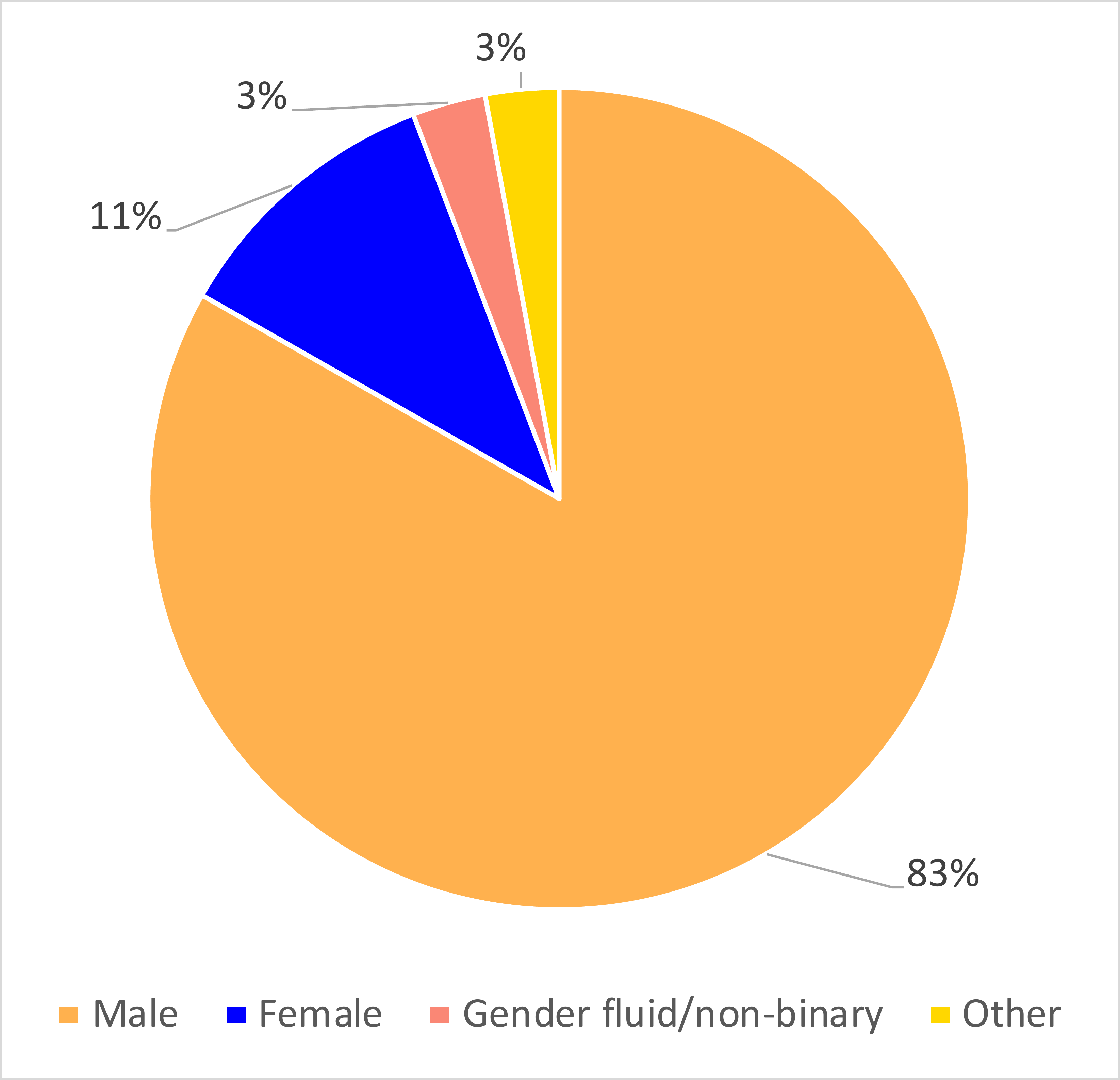}
\hspace{.8in}
\includegraphics[width=2.5in]{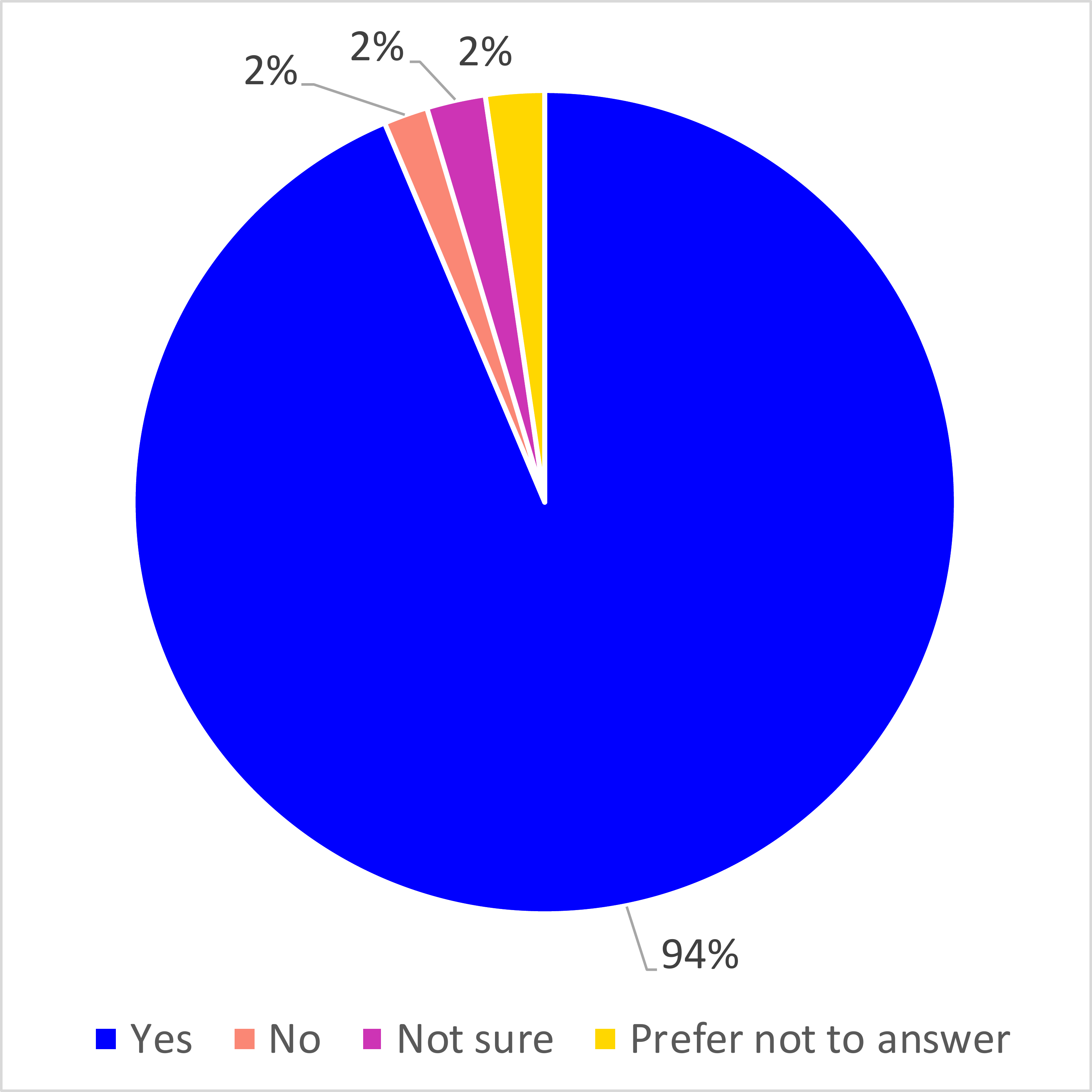}

(a)\hspace{3in}(b)

\includegraphics[width=2.5in]{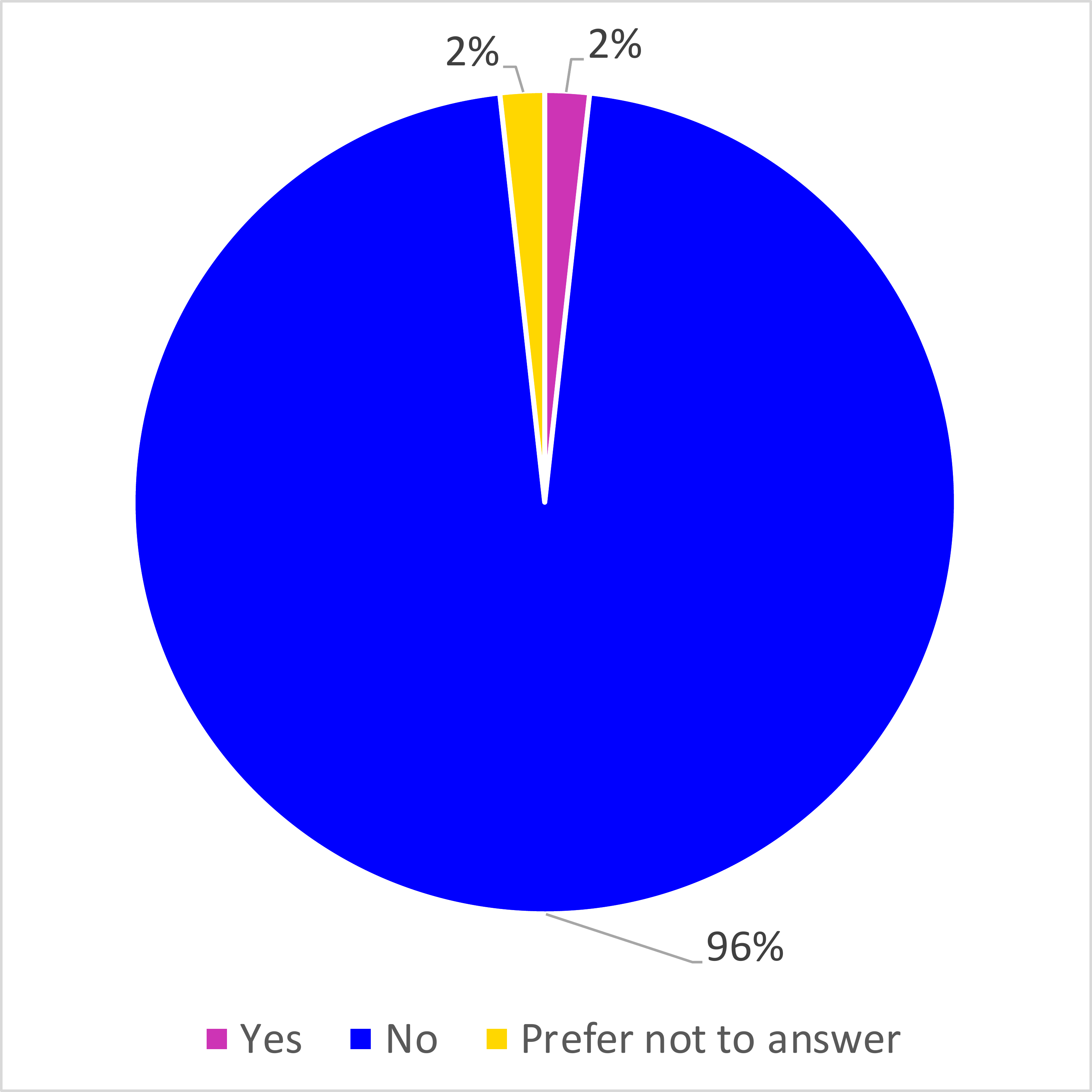}

(c)

\caption{(a) What best describes your gender identity?
(b) Do you think most people are aware of your 
gender identity?
(c) Do you identify as transgender?}
\label{fig:genderID}
\end{center}
\end{figure}

\begin{figure}[htbp]
\begin{center}
\includegraphics[width=2.5in]{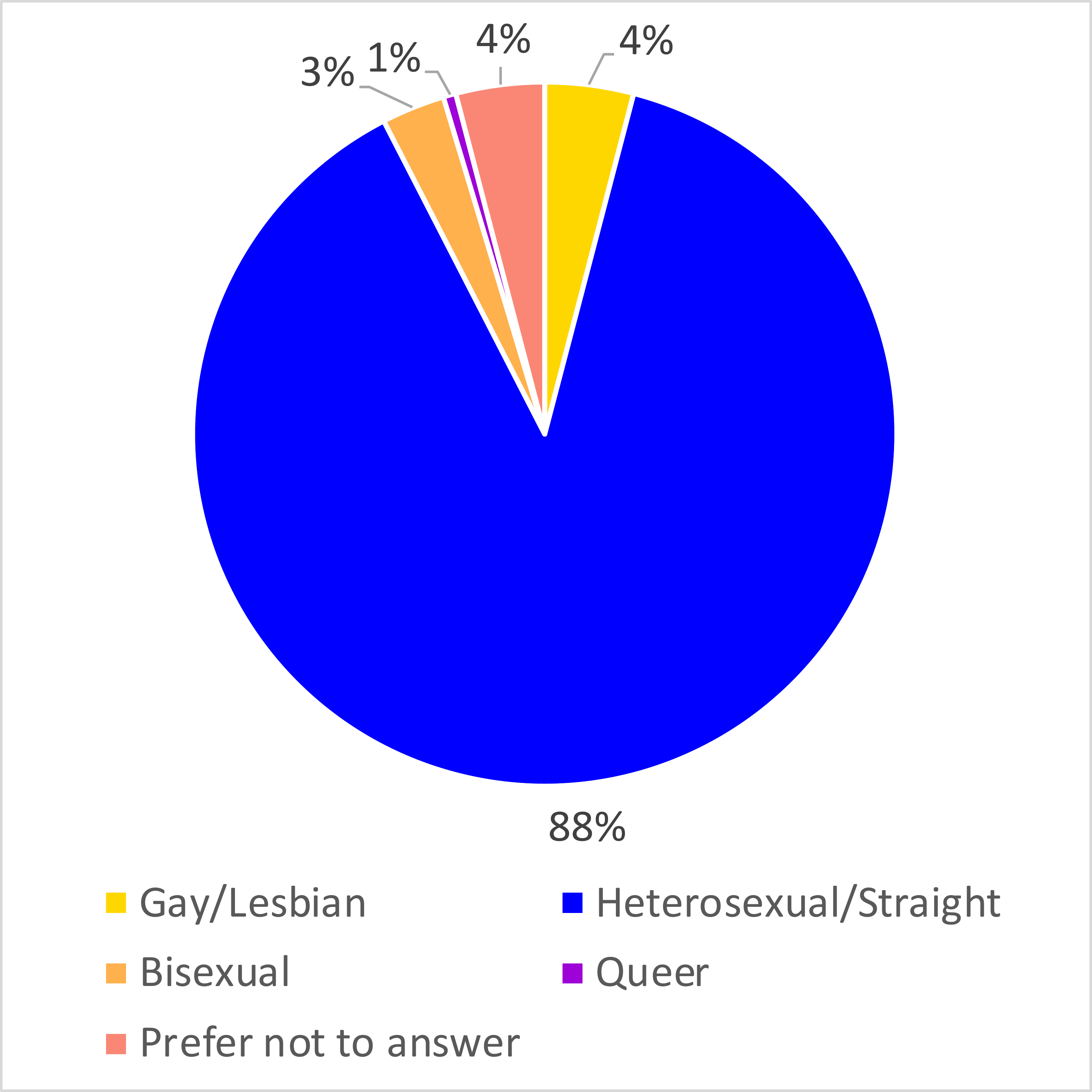}
\hspace{.8in}
\includegraphics[width=2.5in]{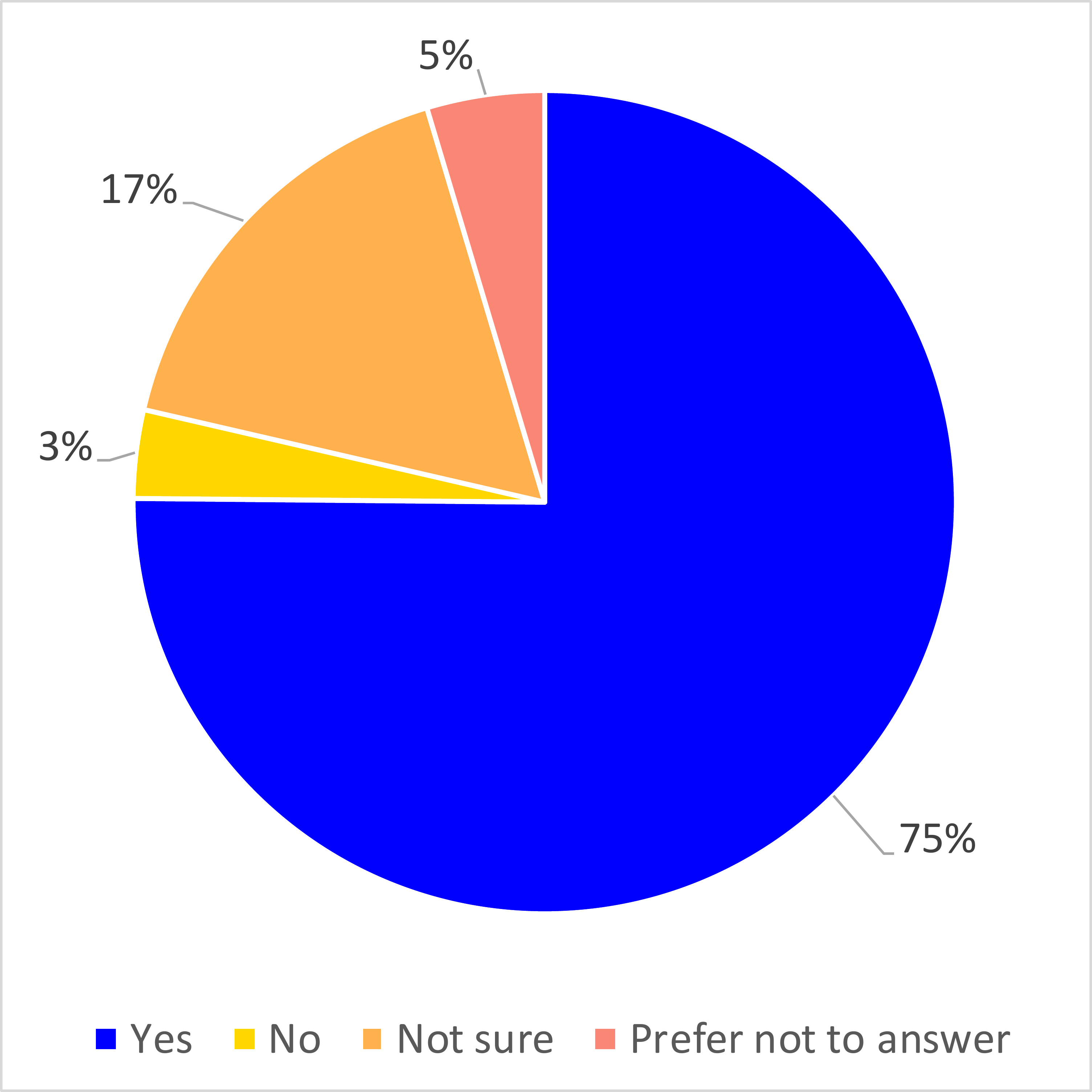}

(a)\hspace{3in}(b)

\caption{(a) What is your sexual orientation?
(b) Do you think most people are aware of your 
sexual orientation?}
\label{fig:sexualorientation}
\end{center}
\end{figure}

\newpage

\section*{Appendix 2: Women in Lattice Statistics (Updated)}\label{app2}

\begin{figure}[htbp]
\begin{center}
\includegraphics[width=3.4in]{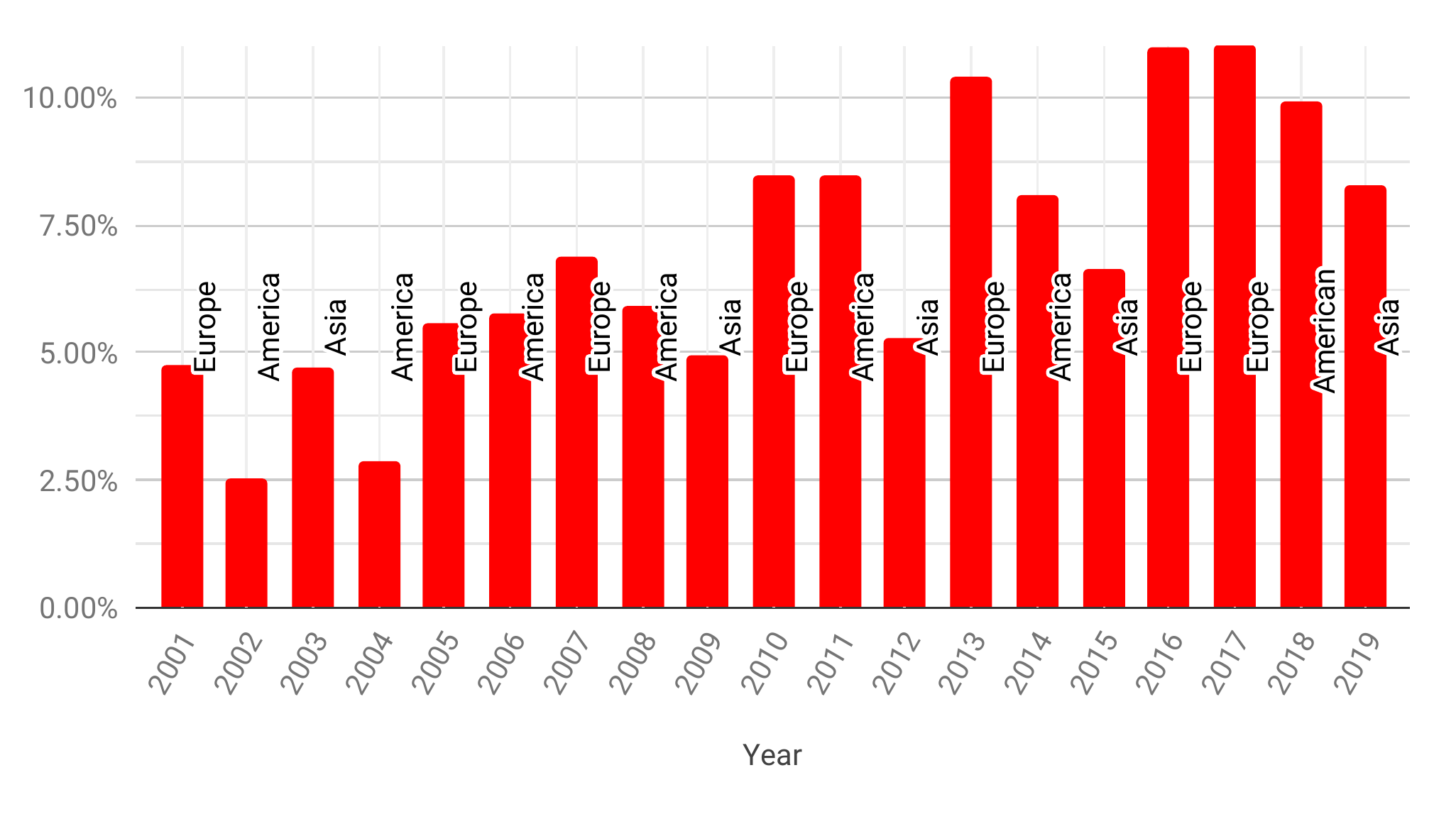}
\includegraphics[width=2.4in]{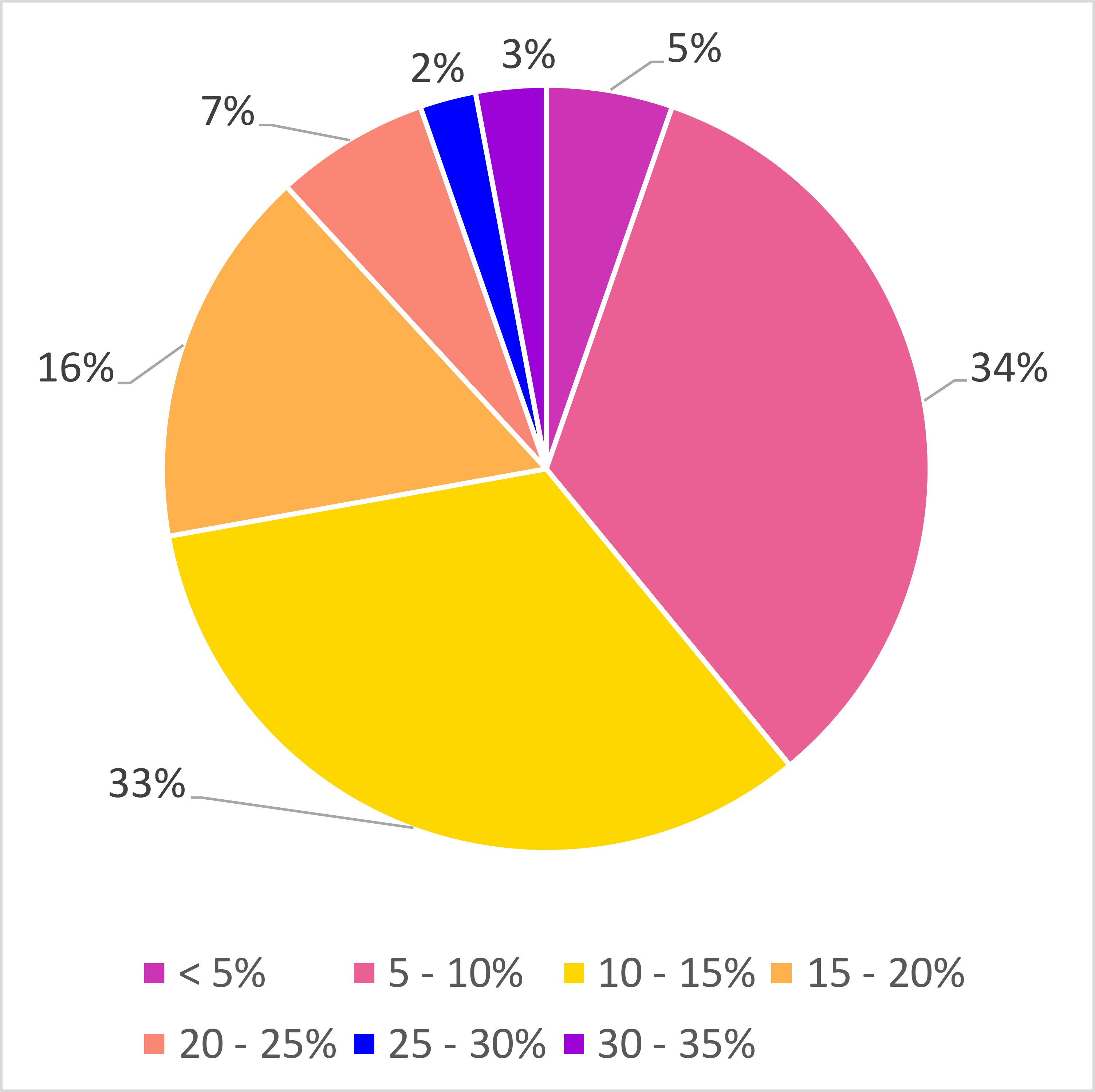}

(a)\hspace{3in}(b)

\caption{(a) Women fraction of lattice conference participation, compared
with (b) the fraction of women perceived to be in attendance by 
respondents to the LDIC survey.}
\label{fig:fractionwomen}
\end{center}
\end{figure}

The fraction of female participants has increased 
relative to the previous decade, as can be seen in 
Fig.~\ref{fig:fractionwomen}. This is likely correlated with a 
global trend of increasing PhD-degrees obtained 
by female students. 
We can compare this with the question on the LDIC survey
which asked respondents to estimate the percentage of women
at the last lattice conference they attended. About one-third
of respondents guessed in the 5-10\% range, which was true
in most of the conferences in the last ten years. About one-third
assumed there were 10-15\% female participation which is only
true in three conferences (2013, 2016, and 2017). 28\% of
survey participants chose a response that was >15\%, showing
that a significant number of participants easily 
overestimate the number of female participants at the conference.

It 
would be interesting to cross-check the observed trend with 
the recent APS data.
Conferences in the EU continue to have larger fractions 
of female participation. Are there lessons that US/Asia can learn 
to improve the female participant rates or is this due 
to culture/grants limitations?

\begin{figure}[htbp]
\begin{center}
\includegraphics[width=4in]{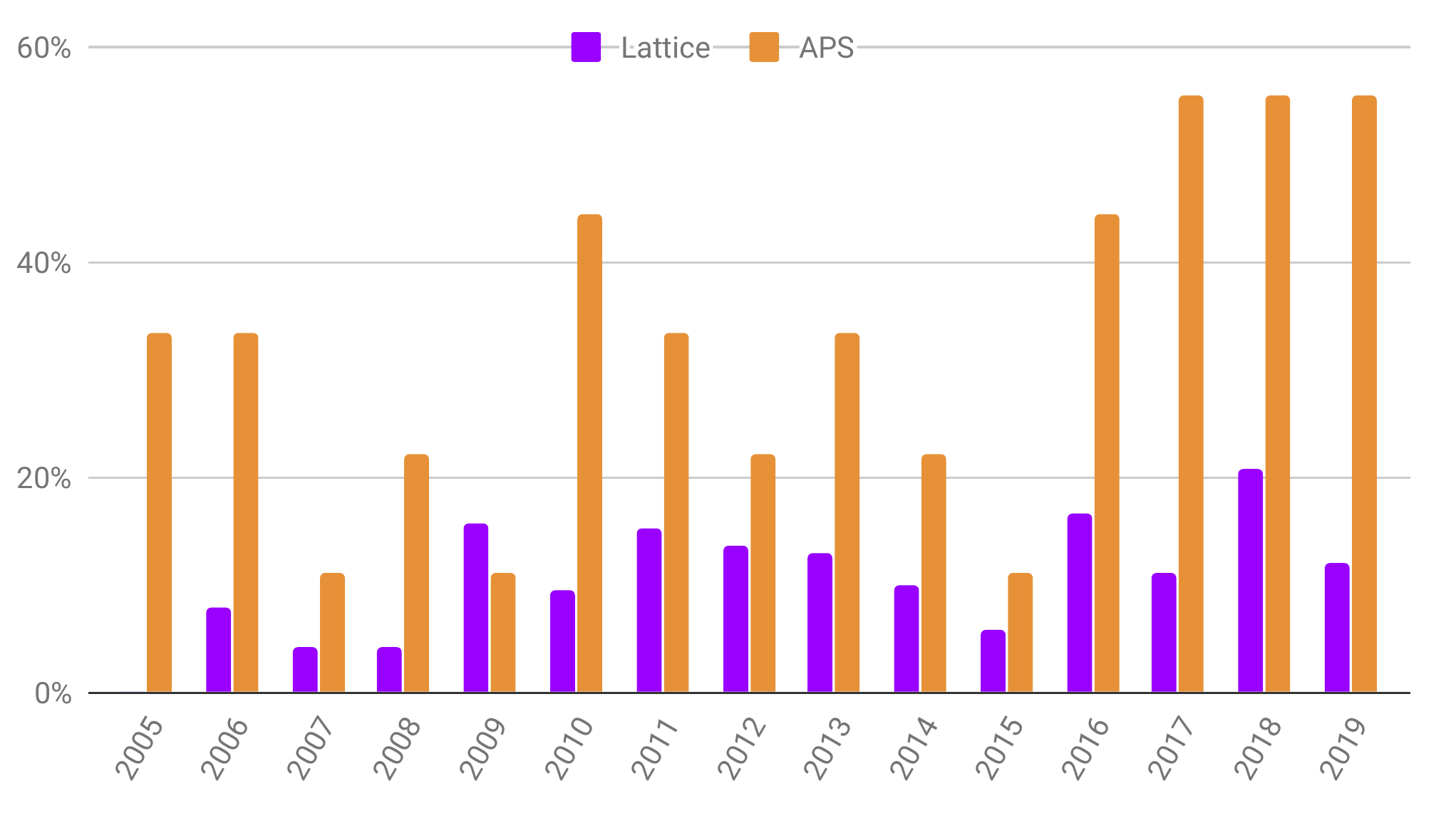}
\caption{Comparison of the fraction of women plenary speakers between
lattice conferences and the April APS meetings.}
\label{fig:LQCDvsApril}
\end{center}
\end{figure}

In Fig.~\ref{fig:LQCDvsApril} we show the percentage 
of female plenary speakers at the 
lattice conference (purple) and the APS April Meeting (orange) as 
a function of year. Most of lattice conference has around 10\% 
of female speakers. In comparison with APS April meeting, which 
covers similar fields in nuclear and particle physics, there is room 
for much improvement in Lattice QCD.

\begin{figure}[htbp]
\begin{center}
\includegraphics[width=4in]{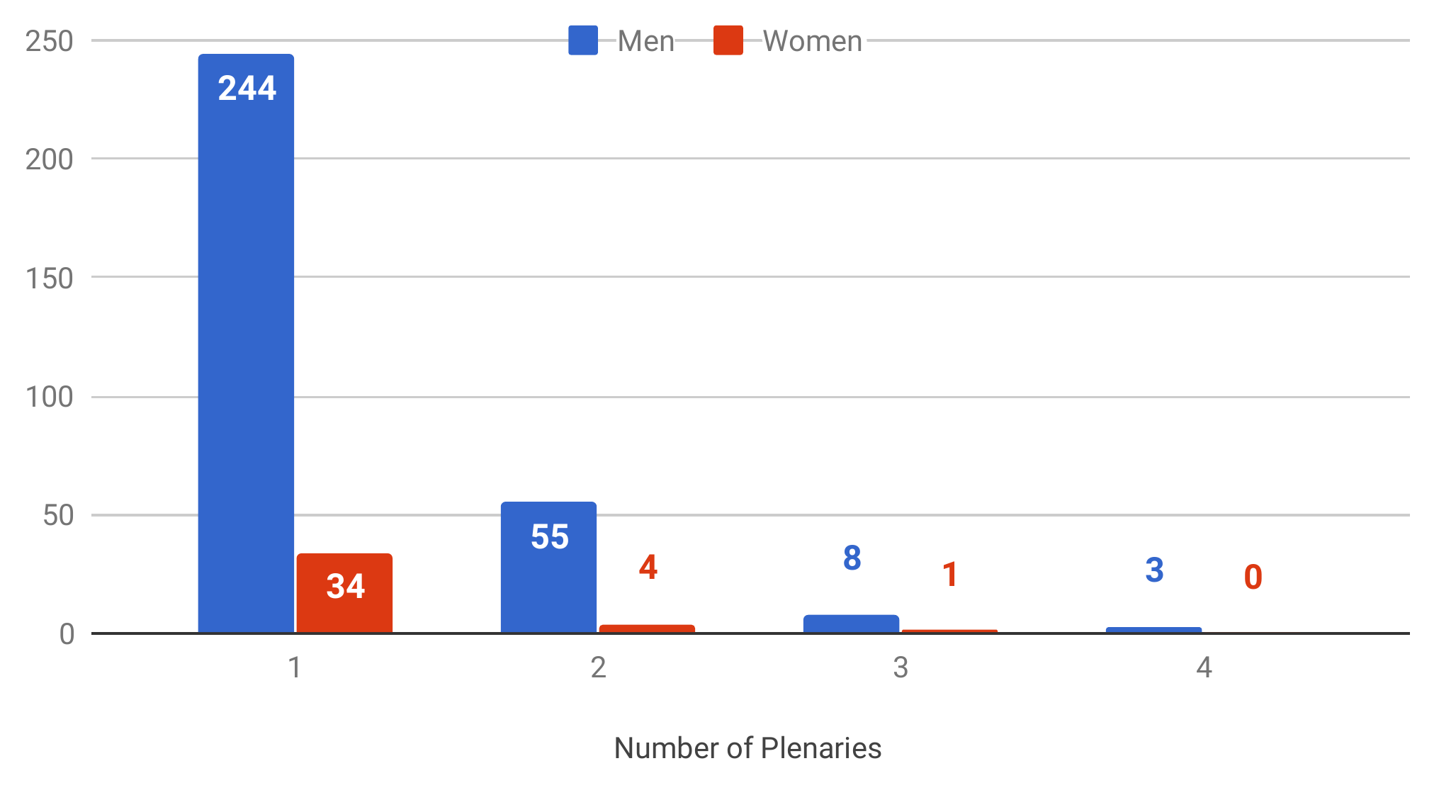}
\caption{Lattice plenary reinvitations by gender.}
\label{fig:reinvites}
\end{center}
\end{figure}

Finally we show the number of male and female plenary speakers as a function 
of the number of times those speakers were invited in 
Fig.~\ref{fig:reinvites}. There is 
a roughly 10\% fraction of women in our field (in contrast with 
APS: $\sim$30\%). Can the small number of plenary talks given by 
women be due to lack of women in our field? There is a 
considerably stronger tendency to re-invite male speakers
for plenary talks, rather than to re-invite past female speakers.

More updated statistics can be found online 
in \href{https://drive.google.com/drive/u/1/folders/1E20ejAj6py4KUJdILoj7J1rPvMtRcsDz}{this google directory}.

\bibliographystyle{JHEP}   

\bibliography{refs}

\end{document}